\documentclass[11pt,a4paper]{article}
\usepackage{natbib}
\usepackage{graphicx}
\usepackage{amsmath}
\usepackage{epsfig}
\usepackage{makeidx}
\usepackage{latexsym}
\usepackage{amssymb}
\usepackage{color}

\begin{document}




\title{On the elastic moduli of three-dimensional assemblies of spheres:
  characterization and modeling of fluctuations in the particle displacement and rotation}


\author{I. Agnolin, J.-N. Roux\\
\small{Laboratoire des Mat\'eriaux et des Structures du G\'enie Civil}\footnote{
LMSGC is a joint laboratory depending on Laboratoire Central des Ponts et Chauss\'ees, \'Ecole Nationale
des Ponts et Chauss\'ees and Centre National de la Recherche Scientifique}\\
\small{Institut Navier, 2 all\'ee Kepler, Cit\'e Descartes, 77420 Champs-sur-Marne, France}}

\date{}
\maketitle


\begin{abstract}
The elastic moduli of four numerical random isotropic packings of
Hertzian spheres are studied. The four samples are assembled with different preparation
procedures, two of which aim to reproduce experimental compaction by vibration and lubrication. The mechanical properties of the samples are found to change with the preparation
history, and to depend much more on coordination number than on density. 

Secondly, the fluctuations in the particle displacements from the average strain are analysed, and the way they affect the macroscopic behavior analyzed. It is found that only the average over
equally oriented contacts of the
relative displacement these fluctuations induce is relevant at the
macroscopic scale. This average depends on coordination number, average geometry of the contact
network and average contact stiffness. As far as the separate contributions from particle displacements and rotations are concerned, the former is found to counteract the average strain along the contact normal,
while the latter do in the
tangential plane. Conversely, the tangential components of the center
displacements mainly arise to enforce local equilibrium, and have a small, and generally stiffening effect at the macro-scale. 

Finally, the fluctuations and the shear modulus that result from two approaches
available in the literature are estimated numerically. These approaches are both based on the equilibrium of a small-sized
representative assembly. The improvement of these estimate with respect to the average strain assumption indicates that
the fluctuations relevant to the macroscopic behavior occur with
short correlation length.
\end{abstract}




\section{Introduction}
Granular media are relevant to civil engineering in
the form of soils or granulates; to agriculture as seeds; to industry
as powders to compact or mix, or because they represent a convenient way to
shape materials that are to store or transport. Granular media are made of grains that interact at contact
points, where forces arise to oppose the relative displacement
between contacting particles. The focus in this work is on the stress-strain relationship for dense assemblies in the linear
elastic regime. This
concerns the reversible response to small disturbances, with the additional constraint that any change
on the geometry of the contact network be negligible. Experimentally, this range is investigated by wave propagation, or by biaxial tests when the technical
apparatus allows to measure very
small deformations, as in Thomann and Hryciw (1990) and Hicher (1996). 

The response of granular media crucially depends on the
structure of the contact network. This is defined by
the average number of contacts per particle and
the texture fabric, and depends on the procedure employed to assemble the sample. Current assembling procedures consist in shakes, taps,
vibration, lubrication or undulatory shear, as in Nicolas et al.
(2000), Jia and Mills (2001) or Cundall et al. (1989). In two-dimensions
the particle fabric is accessible, classically through experiments on
photoelastic disks like those in Majmudare and Behringer (2005) and Atman et
al.) 2005. Photoelastic disks allow detection of contacts through that of colored fringes on the surface of
interacting grains, and computation of the contact forces through
image analysis. On the contrary, the unambiguous identification of the contact
network in three dimensions still represents an unsolved task, due to the
difficulty to distinguish between infinitely close and contacting
grains. Even when advanced
techniques are used, like X-Ray tomography in Aste et al. (2005), the
uncertainty in the determination of the number of neighbors remains significant.
As a consequence, the
assemblies are still customarily characterized in
terms of density, which is the only easily-measured quantity related to the
microstructure. Density still plays a central role in the studies about
granular media, especially in the context of compaction (for example, Procopio and Zavanglios,
2005) or in Cam-Clay-like models (Roscoe and Burland, 1968). Numerical simulations are at present the only tool to access the internal structure. Discrete Element Method (DEM) simulations are currently employed, 
which predict the evolution of a granular assembly by integration in time
of the equations of motion of the individual particles (Cundall and Strack, 1979). These
are treated as rigid bodies that interact at contact points, thereby
neglecting the deformation
of the grains as continua and the mutual influence between close contacts on the
same grain. Such an assumption is reasonable in three dimensions, where the elastic deformation induced by the
force between contacting bodies decays as the square of the distance from the contact
surface (Johnson, 1985). In two dimensions, the decay is only proportional to
this distance. In
Procopio and Zavaliangos (2005)
two-dimensional assemblies of disordered glass disks are simulated by using both a DEM code and a Finite Element Method (FEM) mesh for the individual
particles. Comparison between the two simulations shows that incorporating the behavior of the beads as continua does soften the assembly. 
Like in experiments, the first operation in a numerical simulation is
sample compaction. In the scientific coomunity, this is almost exclusively performed by means of a frictionless
compression, which is easy to reproduce and allows to match experimental
densities. A recent study by Makse et al. (2004) suggests that
the behavior of numerical samples assembled in this way is in agreement with that of real assemblies of glass beads under large pressures, between 10
and 100 MPa. However, to our knowledge neither alternative procedures nor the sample behavior in a smaller pressure range, though recurrent in geotechnical tests, have been sufficiently studied.

In Continuum Mechanics, the interest is in the prediction of the elastic
moduli. A classic attempt
follows from assuming that the relative displacements
between grains are those induced by the average strain. In this way, Digby (1981) and Walton (1987) analytically
infer the bulk and shear modulus of a random assembly of spheres. Comparison
with the elastic moduli of numerical assemblies, as in Cundall et al. (1989),
reveals that the prediction of the bulk modulus is sufficiently accurate, while
that of the shear modulus is severely overestimated. The average strain assumption fails because it does not
incorporate considerations of equilibrium, which is attained by the grains at
the expenses of additional displacement fluctuations in the presence of disorder. In fact, forces and torques
induced by the average strain identically balance only in perfectly ordered packings. This apparently conflicts with the experiments
in Duffy and Mindlin (1957) on regular packs of steel spheres, which still
exhibit a shear modulus different from the average strain prediction, especially at low
pressure. Goddard (1990) proposes as an explanation the squeezing of conical
asperities to Herztian contacts and the buckling of compressed columns, which
are mechanisms that imply a significative evolution of the contact
network. However, the work of Velicky and Caroli (1999) on
hexagonal two-dimensional assemblies of spheres shows that even small dispersions in the
diameter induce important deviations from the average strain prediction.  

Several analytical solutions in the literature attempt to
overcome the insufficiency of the average strain assumption. Some focus on contact forces. For example, Chang et al. (1995)
estimate the elastic moduli of assemblies of frictional spheres whose contact forces
are made depend on the average fabric. They show that the moduli depend on the ratio
between tangential and normal contact stiffness and on the geometry of the
fabric tensor. The same is done in Trentadue (2003),
but in the more general context of non-spherical particles. Other
approaches focus on displacements, solving for them the balance of force and
torque on a small-sized subassembly. This is the case in the Pair Fluctuation (PF) method in
Jenkins and La Ragione (2001), and in the 1-Fluctuating Particle (1FP) method in Kruyt and Rothenburg (2002). These methods have been applied to estimate, in turn, the
shear modulus of dense assemblies of frictionless
spheres and of non-rotating frictional disks. With respect to the average strain assumption, the ratio of the estimate to
the effective value is reduced from 40 to 5 in the first case, and gets close
to 1 in the second case.

The performance of a prediction depends on how realistic its
assumptions are with respect to experimental observations. In two dimensions, the numerical simulations of Calvetti
and Emeriault (1999) and Kruyt and Rothenburg (2001) show that the average force over
equally oriented contacts obeys the directional dependence predicted by the
average strain assumption. That is, isotropic deformations induce isotropic
fluctuations on average, while a biaxial compression induces radial
fluctuations that are largest in the direction of the maximal
compression and tangential ones that are largest at 45 degrees from the principal strains. The two-dimensional simulation in Kuhn
(1999) reveals the appearance of deformation patterns. Gaspar and Koenders
(2001) reexamine this simulation and relate the width of the deformation patterns to the
correlation length of displacement fluctuations, which is found to be of about five
particle diameters. Agnolin and Kruyt (2006) examine the linear elastic
behavior of frictional assemblies of disks of coordination number between 3.5 and 5, and
estimate their shear modulus in the hypothesis that the displacement fluctuations
correlate over four particle diameters. With respect to the average strain
assumption, they obtain a decrease in the ratio of the estimate to the
effective value from more than 3 to about 1.5 in the less favorable case, which corresponds
to small coordination number. Other simulations report much wider correlation lengths. In the 2D simulation in Williams and
Rege (1997) displacement fluctuations occur in circulations cells that
span a length of about twenty-particles. So large correlation lengths are
likely to be induced by the large
deformation applied and to be further emphasised by the particle softness. In this simulation, the deformation reaches $10\%$ of particle diameter in steps of
$1\%$ of particle diameter, which are values characteristic of peack and post-peak behavior in both experiments on sand and numerical
simulations of glass spheres (Thornton and Antony, 1998). A thorough
description of the fluctuations from the average strain from experiments or
numerical simulations and of the way they determine the
macroscopic behavior is still missing in
three dimensions.

In this work, a series of numerical
isotropic assemblies of glass beads is first created. The beads interact through Hertzian contact forces,
whose constitutive law results in an increase of the particle stiffness with
the overlap. Each sample is compacted following a different preparation
procedure, and among the procedures, two aim to reproduce experimental compaction. All procedures result in similar densities at the same pressure, but in
significantly different coordination number. The elastic moduli of the samples are
determined over a wide range of pressure, and compared with their estimate on the base of the average strain
assumption. The fluctuations from the average strain
the samples undergo are characterized, and conclusions are derived about the way they 
determine the behavior of the equivalent continuum. In doing so, distinction
is made between the contributions from the various kinematic ingredients, namely
center displacements and rotations, which are inherent in the discrete nature
of granular materials. Finally, two approaches available in the
literature are discussed and compared, which predict fluctuations consistent with our
experimental observations. Our study fixes
the performance attainable by analytical solutions that would be proposed to
these approaches for frictional spheres, and concludes about the size of the smallest domain which needs to be considered to reliably predict the fluctuations relevant to the macroscopic behavior.

\section{Micromechanics}\label{Micromechanics} 
The simulated grains are glass
beads of diameter $D$ and mass $m$, and are treated as rigid bodies that interact through points of contact. A
contact between two grains, say $i$ and $j,$
is defined by its position and the unit vector $\mathbf{n}^{(ij)}$ that
points from the center of grain $i$ to that of grain $j$. In a counterclockwise Cartesian
reference frame of axes $\mathbf{e}_{1}$, $\mathbf{e}_{2}$, $\mathbf{e}_{3}$, 
\begin{equation*}
\mathbf{n}^{(ij)}=\left( \cos\theta ,\text{ }\sin\theta \cos\phi ,\text{ }
\sin\theta \sin\phi \right),
\end{equation*}
where $\theta $ is the polar angle with respect to axis $1$, and $\phi$ is the
azimuthal angle. Let $\mathbf{u}^{(i)}$ and
$\mathbf{u}^{(j)}$ be the displacement of the two grains; $\boldsymbol{\omega}^{(i)}$ and
$\boldsymbol{\omega}^{(j)}$ their rotation; $\mathbf{u}^{(ij)}$
their relative displacement, and $\mathbf{t}^{(ij)}$ the unit vector aligned with the tangential
component $\mathbf{s}^{(ij)}$ of $\mathbf{u}^{(ij)}$; $\mathbf{s}^{(ij)}$ has absolute value $s^{(ij)}$, i.e.
\begin{eqnarray*}
\mathbf{u}^{(ij)}&=&\mathbf{u}^{(i)}-\mathbf{u}^{(j)}+\frac{D}{2}\left[\boldsymbol{\omega}^{(i)}+\boldsymbol{\omega}^{(j)}\right]\times \mathbf{n}^{(ij)},\\
\mathbf{s}^{(ij)}&=&\mathbf{u}^{(ij)}-u_{a}^{(ij)}n_{a}^{(ij)}\mathbf{n}^{(ij)},\\
s^{(ij)}&=&\lvert\mathbf{s}^{(ij)}\rvert,\\
\mathbf{t}^{(ij)}&=&\mathbf{s}^{(ij)}/s^{(ij)},
\end{eqnarray*}
where the symbol $\times$ and the vertical bars $\lvert...\rvert$ denote, in turn, the vector product and the absolute value. Contacting grains interact by means of contact forces. These oppose their relative displacement at the contact point, and have normal and tangential component to the contact of absolute value $F_{N}^{(ij)}$ and $F_{T}^{(ij)}$, respectively. $F_{N}^{(ij)}$ obeys Hertz's law, i.e. 
\begin{equation}
F_{N}^{(ij)}=\frac{Y\sqrt{D}h^{(ij)3/2}}{(1-\nu ^{2})3},  \label{ForzaN}
\end{equation}
where  $Y$ is the Young's modulus of the material the beads are made of,
$\nu$\ is its Poisson's modulus and $h^{(ij)}$ is the contact overlap. No
tensile normal forces are allowed. The increments in $F_{T}^{(ij)}$ obey
\begin{equation}
dF_{T}^{(ij)}=\frac{2(1-\nu )}{2-\nu }\frac{dF_{N}^{(ij)}}{d h^{(ij)}}ds^{(ij)},
\label{dFT}
\end{equation}
which is Mindlin's (Mindlin
and Deresiewicz, 1953) initial tangential stiffness between two spheres kept in contact by the
normal force $F_{N}^{(ij)}$. $F_{T}^{(ij)}$ is
bounded above by the coefficient of friction $\mu $, so that $|F^{T}|\leq \mu
|F^{N}|$. Along some loading-unloading paths (Elata
and Berryman, 1996), use of (\ref{dFT}) might induce
spurious creation of energy, which is avoided by incorporating the recommendations
by Johnson and Norris (1996). The contact force $\mathbf{F}^{(ij)}$ of grain $j$ on grain $i$ can thus be written as
\begin{equation*}
F_{r}^{(ij)}=-F_{N}^{(ij)}n_{r}^{(ij)}-F_{T}^{(ij)}t_{r}^{(ij)}.
\label{contactForce}
\end{equation*}

At the macroscopic level, the assembly is characterised by its solid fraction $\Phi $
and coordination numbers $z$ and $z^{*}$, namely
\begin{eqnarray}
\Phi  &=&\frac{N\pi D^{3}}{6V},  \notag \\
z &=&\frac{2C}{N},  \notag\\
z^{*} &=&\frac{2C}{N_{act}},\label{coord}
\end{eqnarray}
where $N$ is the number of grains in the assembly, $N_{act}$ is that of
force-carrying grains and $C$ is the number of contacts. The average number of contacts per active particle $z^{*}$ emphasises the role of the force-carrying network. One has that $z=z^{*}N_{act}/N$. The macroscopic stress $\sigma _{rm}$ in a numerical assembly (Cundall, 1983) is given by 
\begin{equation}
\sigma _{rm}=-\frac{D}{2V}\sum_{i=1}^{N_{act}}
\sum_{j=1}^{N(i)}F_{r}^{(ij)}n_{m}^{(ij)},  \label{defStress}
\end{equation}
which is the average over the volume $V$ of the assembly of
Cauchy's stress as defined by Love (1944). Its components are taken positive in compression. In (\ref{defStress}), the index $j$ spans the
$N(i)$ neighbors of grain $i$, and the factor $1/2$ avoids twice-counting
contacts.  The components of the strain
tensor $\bold{E}$ measure the relative
change in length and are taken positive in compression. For example,
\begin{equation*}
E_{11}=\left( L_{1}-l_{1}\right) /L_{1},
\end{equation*}
where $L_{1}$ and $l_{1}$ measure the cell length along the direction
1 before and after the deformation. In elastic systems, increments in stress
and strain are related by the fourth-order tensor $\mathbf{C}$\textbf, 
\begin{equation}
\Delta \sigma _{rm}=C_{rmij}E_{ij},
\label{4Ctens}
\end{equation}
where Einstein's convention is used. In isotropic systems, the
tensor $\mathbf{C}$ has two indipendent components, say $C_{1111}$ and
$C_{1122}$. Equivalently, the stress-strain relationship can be expressed in
terms of the bulk and shear modulus $B$ and $G$, as
\begin{eqnarray}
B &=&\left( C_{1111}+2C_{1122}\right) /3, \notag\\
G &=&\left( C_{1111}-C_{1122}\right) /2.
\label{relaz}
\end{eqnarray}

\begin{figure}[h]
{\small \centering
\includegraphics[width=14cm]{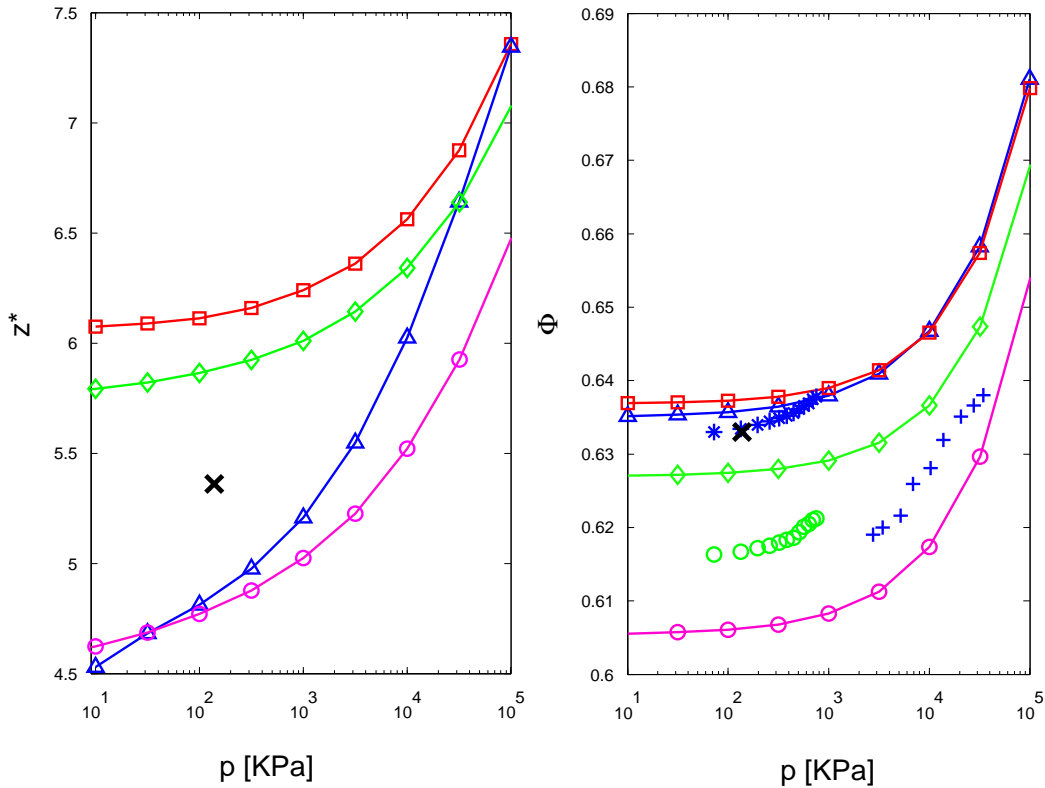}  }
\caption{{Evolution of coordination number $z^{*}$ (average number of contacts per active particle) and density $\Phi $ with
pressure for numerical samples A($\Box $), B$(\Diamond )$, C($\triangle $)
and D($\bigcirc $). Symbols (*) and (o) are, in turn, for vibrated and lubricated
samples in the experiments by Jia and Mills reported in (Agnolin et al., 2005); (+) for the experiments by Domenico
(1977); the isolated thick $\mathbf{x}$ indicates the numerical simulation by
Cundall et al. (1989).}}
\label{zphi}
\end{figure}

\section{DEM simulations}\label{NumericalSim}
\subsection{Creation of the initial state}\label{InitialState}
DEM simulations are employed to create four numerical
assemblies, labelled by $A$, $B$, $C$ and $D$, each made of 4000 glass spheres confined in a periodic
cell. As far as the mechanical properties of glass are concerned, $Y=70\;GPa$ and $\nu=0.3$ are its Young's and Poisson's modulus, respectively. The evolution of an assembly is studied by integrating in time the equations of motion of the individual grains, which interact through contact forces. Each sample is assembled following a different preparation procedure, and is at
equilibrium at the pressure of $10KPa$. The lowest pressure points in Figure \ref{zphi} give
density and coordination number of the initial
states. By ``equilibrium'' it is meant that
the resultant force and torque on each grain are smaller than $pD^{2}/10^{4}$ and 
$pD^{3}/10^{4}$, respectively. Moreover, the kinetic energy has to be smaller than
$pD^{3}/10^{7}$, and once all velocities have been set to zero, the kinetic
energy must stay null. The initial
state of sample A results from the isotropic compression of a frictionless
granular gas up to the pressure of $10KPa$, and has coordination number
close to 6. Procedure $A$ is a classic in the literature
about granular media (Thornton, 2000; Makse et al., 2004), as it is easy to reproduce and gives good agreement
with experimental solid fractions. In the case of sample B, the isotropic
compression is applied on slightly frictional grains $(\mu =0.02),$ as in an
imperfect lubrication. The initial state of sample $B$ is
less dense than that of sample $A$, but still has a large
coordination number. The procedure employed to create sample $C$
interprets vibration, which is currently used to compact
experimental samples, as a temporary suppression of friction. In this
procedure, the coordinates of the grains of the initial state of sample $A$
are first scaled by a factor 1.005. In this way, all contacts are
opened. Then, the grains are mixed with random collisions that preserve
kinetic energy until they undergo 50 collisions on average. Only at this
point is the isotropic compression resumed and continued up to the pressure $
P=10KPa$, with friction $\mu =0.3$. Sample $C$
is only slightly less dense than $A$, but has a
much lower coordination number, because of friction, and $13\%$ of its grains do not carry any force. These grains are
called ``rattlers'' in the literature. The comparison between density and
coordination number of samples $A$ and $B$ on
one side and $C$ on the other emphasizes that large
densities do not necessarily imply large coordination numbers. Lastly, sample 
$D$ is the result of the isotropic compression of a frictional gas with friction
coefficient $\mu =0.3$. The initial
state of sample $D$ has low density and low coordination
number, and $11\%$ of its grains are rattlers.

\subsection{Isotropic compression and elastic moduli}
\label{IsotropicCompression}
From their initial state, the samples are slowly isotropically
compressed up to the pressure $p=10^{5}KPa$ via DEM simulations. By `slow' it is meant that the dimensionless control parameter $I=\dot{E}\sqrt{m/Dp}$ stays smaller than $10^{-4}$. In the expression of $I$, $E$ is the
absolute value of the macroscopic deformation. The parameter $I$ has been defined in (da Cruz et al., 2005), and compares
the macroscopic deformation rate to the particle-scale time necessary for a
grain initially at rest to cover the distance $D/2$ when pushed by a force $
F=pD^{2}$.

\begin{figure}[t]
{\small \centering
\includegraphics[width=14cm]{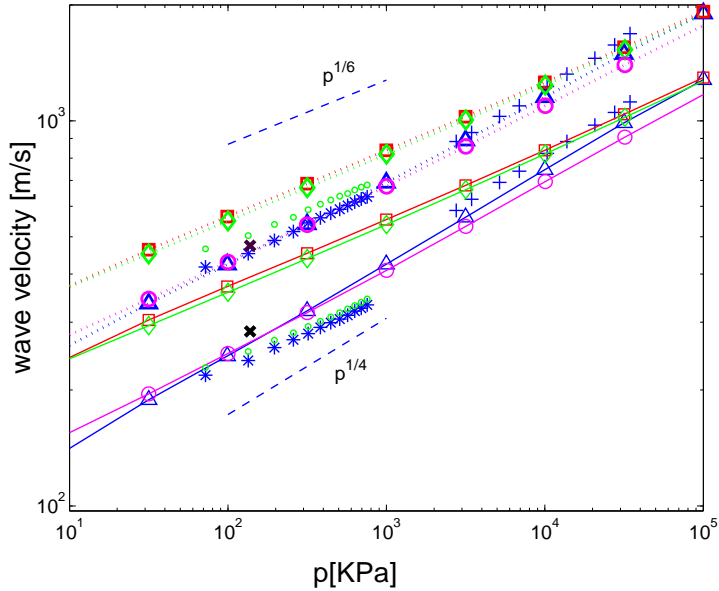}  }
\caption{{Wave velocities for numerical samples A($\Box $), B$(\Diamond )$,
C($\triangle $) and D($\bigcirc $). Dotted and solid lines are used for
longitudinal and shear wave velocity $v_{l}$ and $v_{s}$, respectively.
Symbols * and $\circ $ are used, in turn, for the experimental results on dry
and lubricated grains in Agnolin et al.
(2005); + for those in
Domenico (1977); the isolated thick $x$ indicates the numerical simulation by
Cundall et al. (1989). For each experiment, the upper and lower series
measure, in turn, $v_{l}$
and $v_{s}$.}}
\label{Vel}
\end{figure}

At eight intermediate values of pressure the assemblies are allowed to reach equilibrium, and their bulk and shear modulus $B$ and $G$ are measured. $B$ is determined by measuring the
macroscopic strain $E_{rm}$ induced by a small
isotropic increment in stress 
\begin{equation*}
\Delta \sigma_{rm}=\Delta p\;\delta_{rm},
\end{equation*}
where $\delta_{rm}$ is the identity tensor. Due to (\ref{4Ctens}) and (\ref{relaz}),
\begin{eqnarray}
B &=&\Delta p/3E_{ii}. \notag  \label{B} 
\end{eqnarray}
The shear modulus $G$ is determined by measuring the macroscopic strain induced by an incremental stress of the form
\begin{equation}
\Delta \mathbf{\sigma }=\Delta q\left[ 
\begin{array}{ccc}
1 & 0 & 0 \\ 
0 & -1/2 & 0 \\ 
0 & 0 & -1/2
\end{array}
\right],
\label{shear}
\end{equation}
as
\begin{eqnarray}
G &=&\Delta q/2\left( E_{11}-E_{22}\right).  \label{modG}
\end{eqnarray}
The elastic
moduli are plotted in
Figure \ref{Vel} in terms of longitudinal and transversal wave velocities $v_{L}$\ and $v_{S}$, 
\begin{eqnarray*}
v_{L} &=&\sqrt{(B+4G/3)/\rho },  \notag \\
v_{S} &=&\sqrt{G/\rho},\label{VelBG}
\end{eqnarray*}
where $\rho $ is the density of the assembly. The elastic moduli of samples C and D are clearly separated from
those of samples A and B, each pair having similar coordination
number. Numerical compaction by vibration gives denser
but less stiff samples than lubrication, in accordance with experiments, but in contrast with the
general belief that the larger the density the stiffer the assembly.

The plot of wave velocities allows comparison with wave propagation
experiments. We consider the experiments reported in (Agnolin et al., 2005) and (Domenico,
1977) on samples of glass beads, whose density evolution with pressure is
also shown in Figure
\ref{zphi}. Only sample C matches the experiments satisfactorily in terms of
both stiffness and density, which makes its study relevant. On the contrary, sample A, whose properties are
generally referred to in the literature, is too
stiff in the range of pressures usual in geotechnical tests. This completes
the observations by Makse et al. (2004), who compare sample $A$
only to the large
pressure tests of Domenico and conclude that it is
representative of real assemblies. In the range of pressure considered by Domenico, the
configurations of samples $C$
converge to those of sample $A$, as the former is
derived from the latter. In both
experiments and numerical simulations, compaction by vibration gives denser
but less stiff samples than lubrication. Also shown in Figures \ref
{zphi} and \ref{Vel}\ are density, coordination number and mechanical
properties of the numerical sample studied in Cundall et al. (1989). This
numerical simulation aims to reproduce the mechanical behavior of a real
assembly created by raining spherical beads into a membrane and tamping
them. The numerical sample is created by means of an initial almost frictionless
compression followed by a series of increasingly frictional expansions and
compressions, up to the point where experimental pressure and density are matched. The
shear modulus of the numerical sample is 127 MPa, which compares well with
the experimental measurement of 161 MPa. Its coordination number is smaller
than 6, as in the case of sample C. All these observations emphasize a much stronger dependence of the elastic moduli
on coordination number than on density. They also show that some of the numerical samples exhibit properties of experimental ones. A thorough analysis of the resemblance between numerical simulations and experiments requires
an extensive study of the internal structure and comparison with the recent
experiments reported in Aste et al. (2005). These are deferred to a publication to appear (Agnolin and Roux, 2007), as the present article focuses on the elastic moduli of the numericla samples. In the referred to
publication, the reader will also find an accurate discussion of the numerical
procedures employed.

\subsection{Linear elastic regime}\label{LER}
In the previous section, the elastic moduli of the numerical samples have been measured by employing DEM simulations. In order for elasticity to apply, these have to be free from dissipation and quasi-static, while linearity requires that all change in geometry be negligible at the first order. This paragraph identifies the range of deformation inside which such requirements are satisfied, and thus validates the results of the previous section.

Quasi-staticity implies that in the transition between
  two equilibrium states all dynamic contributions stay
  negligible. This reduces the forces on the
  grains to their Hertzian interaction, whose expression can be finally linearized for
  small enough deformations. An increment $\Delta F_{r}^{(ij)}$ in the contact force that grain $j$ exerts on grain $i$ can thus be related to the relative
displacement ${u}^{(ij)}$ between the two grains via a normal and a tangential contact stiffness $K_{N}^{(ij)}$ and $K_{T}^{(ij)}$. From (\ref{ForzaN}), the normal contact stiffness
has the form
\begin{equation}
K_{N}^{(ij)}=\frac{dF_{N}}{dh^{(ij)}}=\frac{Y\sqrt{Dh^{(ij)}}}{2(1-\nu
^{2})},  \label{incremKN}
\end{equation}
while (\ref{dFT}) still holds along the tangential direction. $K_{N}^{(ij)}$
and $K_{T}^{(ij)}$\ define the contact stiffness operator 
\begin{equation}
K_{ra}^{(ij)}\doteq K_{N}^{(ij)}n_{r}^{(ij)}n_{a}^{(ij)}+K_{T}^{(ij)}(\delta
_{ra}-n_{r}^{(ij)}n_{a}^{(ij)}),  \label{StiffOper}
\end{equation}
which makes it possible to write that
\begin{equation}
\Delta F_{r}^{(ij)}=-K_{ra}^{(ij)}u_{a}^{(ij)}.  
\label{DForce}
\end{equation}
The relative displacement ${u}_{a}^{(ij)}$ at the contact point
between grains $i$ and $j$ is the sum of the
contributions $u_{a}^{E(ij)}$ from the average strain and $\tilde{u_{a}}
^{(ij)}$ from fluctuations. If $\tilde{u}_{a}^{(i)}$ is the fluctuation of grain $i$ in the center displacement and $\tilde{\omega}
_{a}^{(i)}$ that in the rotation about it, and $\tilde{u}_{a}^{(j)}$ and $\tilde{\omega}
_{a}^{(j)}$ are the analogous quantities for grain $j$, 
\begin{eqnarray}
&&u_{a}^{(ij)}=u_{a}^{E(ij)}+\tilde{u}_{a}^{(ij)},  \notag \\
&&u_{a}^{E(ij)}=DE_{am}n_{m}^{(ij)},  \notag \\
&&\tilde{u}_{a}^{(ij)}=\tilde{u}_{a}^{(i)}-\tilde{u}_{a}^{(j)}+\frac{D}{2}
\epsilon _{abc}\left[ \tilde{\omega}_{b}^{(i)}+\tilde{\omega}_{b}^{(j)}
\right] n_{c}^{(ij)},  
\label{displ}
\end{eqnarray}
where $\epsilon _{abc}$ is Edington's
epsilon. The balance of force and torque on grain $i$ requires that
\begin{eqnarray}
&&\sum_{j=1}^{N(i)}\Delta F_{r}^{(ij)}=0,\notag \\
&&\sum_{j=1}^{N(i)}\epsilon _{abr}n_{b}^{(ij)}\Delta
F_{r}^{(ij)}=0,
\label{eqF}
\end{eqnarray}
where the index $j$ spans over the $N(i)$ neighbors of $i$. Lastly, if the increment
in stress $\Delta \sigma _{rm}$ applied in the DEM simulation is to balance,
\begin{equation}
-\frac{D}{2V}\sum_{i=1}^{N_{act}}\sum_{j=1}^{N(i)}\Delta
F_{r}^{(ij)}n_{m}^{(ij)}=\Delta \sigma _{rm}.  
\label{eqS}
\end{equation}
According to the definition of linearity, the
components of the normal contact vector in (\ref{eqF}) and (\ref{eqS}) are taken to be those found before the increment
in stress is applied. Introducing (\ref{DForce}) and (\ref{displ}) in the system of (\ref{eqF}) and
(\ref{eqS}) makes it solvable for the 3 non-zero components of the strain and
the $6N_{act}$ grain fluctuations. The deformation range inside which the results coincide with those of the DEM simulations strongly depends on coordination number and increases with pressure. For confining pressure between $10KPa$ and $1MPa$, it varies between $10^{-6}$ and $10^{-5}$ for sample $C$, and between $10^{-5}$ and $10^{-4}$ for sample $A$. For larger deformations grain rearrangement
cannot be neglected. The fluctuations that result from these calculations will be analysed in section \ref{SectionFluct}.

\begin{figure}[t]
\begin{center}
{\small \includegraphics[width=12cm]{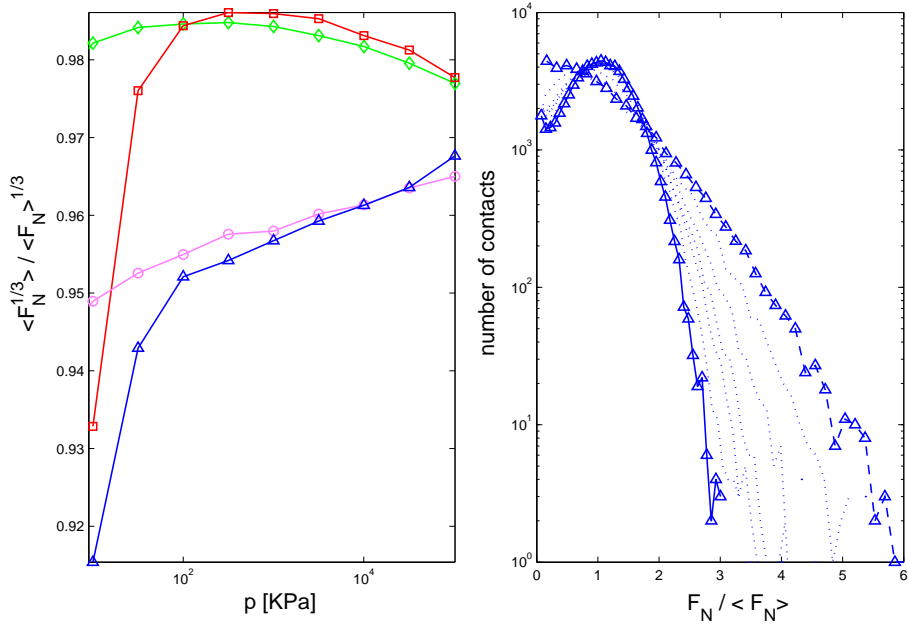}  }
\end{center}
\caption{{\protect\footnotesize { Left-hand side: evolution of the quantity $\langle F^{1/3}\rangle /
\langle F\rangle ^{1/3}$ for samples A($\Box $), B$(\Diamond )$, C($
\triangle $) and D($\bigcirc $). Right-hand side: evolution of the normal
contact force distribution for sample C; the dashed and solid line reperesent,
in turn, the distribution at $p=10$ and $10^{5}KPa$; dotted lines are used at intermediate pressures.}}}
\label{Distr13}
\end{figure}

\section{Average strain assumption}\label{AverageStrain}
\begin{figure}[t]
{\small 
}
\par
\begin{center}
{\small 
\includegraphics[width=14cm]{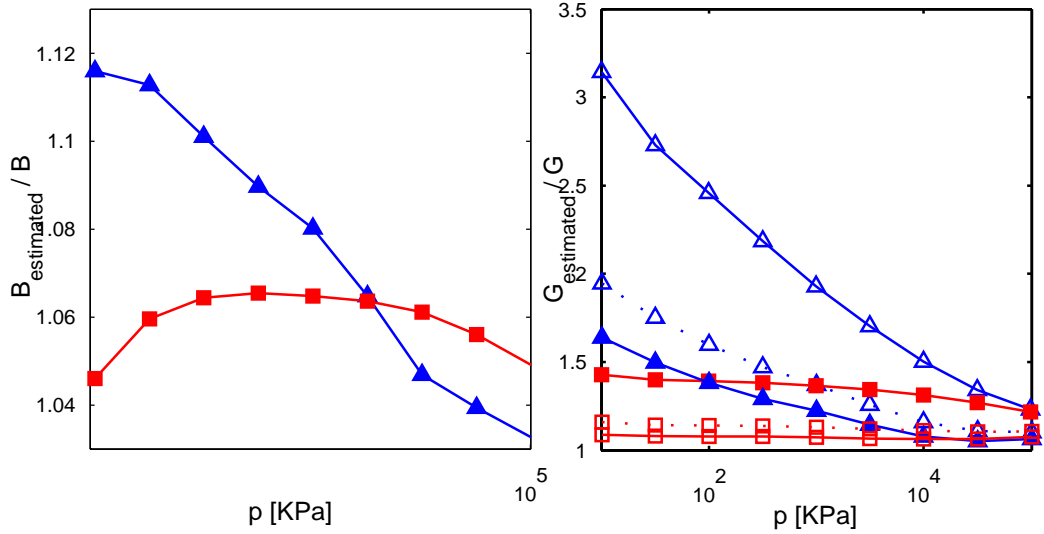}}
\end{center}
\caption{{Ratio of estimated to measured bulk (left) and shear (right) modulus
    for sample A($\Box $) and C($\triangle $). Filled symbols indicate the
    results of the average strain assumption; empty symbols indicate the
    estimate of the shear modulus in the 1FP (dashed line) and the FP approach (solid line).}}
\label{Qual}
\end{figure}

In the attempt to predict the elastic moduli, the easiest
scenario possible is that of relative displacements between contacting
grains driven by the average strain. The corresponding incremental overlap $
h^{E(ij)}$ and tangential relative displacement $s_{a}^{E(ij)}$ would be 
\begin{eqnarray}
h^{E(ij)} &=&DE_{lm}n_{l}^{(ij)}n_{m}^{(ij)},  \notag \\
s_{a}^{E(ij)} &=&DE_{am}n_{m}^{(ij)}-h^{E(ij)}n_{a}^{(ij)},  \label{displE}
\end{eqnarray}
which depend on the contact orientation alone. Expressions (\ref{DForce}) and
(\ref{displE}) are introduced in (\ref{defStress}), and the discrete expression for the stress
is transformed into an integral over the solid angle. To this aim, an isotropic probability distribution function
$f(\psi)=z/4\pi$ is employed, which is defined in such a way that $f(\psi)\,d\psi$ is the number of
contacts in the element of solid angle $d\psi$. As a result,
\begin{equation}
\Delta\sigma _{rm}=-\frac{D^{2}N_{act}}{2V}\langle K_{ra}\rangle\int_{0}^{4\pi} E_{ab}n_{b}n_{m}f(\psi)\,d\psi,  \label{ContStress}
\end{equation}
where $K_{ra}$ is the average contact stiffness. From (\ref{dFT}) and (\ref
{incremKN}), 
\begin{eqnarray}
<K_{N}> &=&\left[ \frac{3DY^{2}}{8\left( 1-\nu ^{2}\right) ^{2}}\right]
^{1/3}<F_{N}^{1/3}>, \notag \\
<K_{T}> &=&\frac{2(1-\nu )}{2-\nu }<K_{N}>,
\label{AvK}
\end{eqnarray}
which with (\ref{4Ctens}) and (\ref{relaz}) give the estimates 
\begin{eqnarray}
B^{E} &=&\frac{1}{2}\left[ \frac{z\Phi Y}{3\pi }\right] ^{2/3}P^{1/3}\frac{
<F_{N}^{1/3}>}{<F_{N}>^{1/3}},  \notag \\
G^{E} &=&\frac{6+9\langle K_{T}\rangle /\langle K_{N}\rangle }{10}B^{E}.
\label{BEGE2}
\end{eqnarray}
For contacts between glass spheres, $\langle K_{T}\rangle /\langle
K_{N}\rangle=0.823$. Expressions (\ref{BEGE2}) account for the width of the normal force
distribution, whose evolution with pressure for sample $C$ is shown in Figure \ref{Distr13}. If this is assumed uniform, the estimates reduce to those of Walton (1987):
\begin{eqnarray}
B^{EW} &=&\frac{1}{2}\left[ \frac{z\Phi Y}{3\pi }\right] ^{2/3}P^{1/3}, 
\notag \\
G^{EW} &=&\frac{6+9\langle K_{T}\rangle /\langle K_{N}\rangle }{10}B^{EW}.
\label{BEGEW}
\end{eqnarray}
Figure \ref{Distr13} shows that the difference between estimates (\ref
{BEGE2}) and (\ref{BEGEW}) is rather small, and larger for samples $A$ and $C$. The ratio of $B^{E}$ and $G^{E}$
to the effective moduli is plotted in Figure \ref{Qual}, where it is observed
that the prediction of
the bulk modulus is reliable. On the contrary, that of the shear modulus
performs poorly, and improvement can be obtained only at the expenses of
incorporating the displacement fluctuations into the modelling.

\section{Fluctuations}\label{SectionFluct}
We study the displacement fluctuations that arise when the shear stress
(\ref{shear}) is applied, and interpret the role of the various kinematic ingredients. Coherently with the constitutive law for the contact force (\ref{DForce}), the fluctuations are
analyzed in terms of the relative displacement they induce at the contact
points. This is denoted by $\tilde{u}_{a}^{(ij)}$ in (\ref{displ}). Its normal and tangential
components $\tilde{h}^{(ij)}$ and $\tilde{s}_{a}^{(ij)}$ are 
\begin{eqnarray}
\tilde{h}^{(ij)} &=&\tilde{u}_{l}^{(ij)}n_{l}^{(ij)},  \notag \\
\tilde{s}_{a}^{(ij)} &=&\tilde{u}_{a}^{(ij)}-\tilde{h}^{(ij)}n_{a}^{(ij)}.
\label{dtilde}
\end{eqnarray}
In the tangent plane, $\tilde{s}_{a}^{(ij)}$ is still the sum
of $\tilde{w}^{(ij)}t_{a}^{E(ij)}$ and $\tilde{z}_{a}^{(ij)}$, in turn
aligned with and perpendicular to the tangential displacement $s_{a}^{E(ij)}$
induced by the average strain. If $t_{a}^{E(ij)}$ is defined as the unit vector aligned
with $s_{a}^{E(ij)}$, 
\begin{eqnarray}
t_{a}^{E(ij)} &\doteq &{s_{a}^{E(ij)}}/|s^{E(ij)}|,  \notag \\
\tilde{w}^{(ij)} &=&\widetilde{s_{l}}^{(ij)}t_{l}^{E(ij)},  \notag \\
\tilde{z}_{a}^{(ij)} &=&\tilde{s}_{a}^{(ij)}-\tilde{w}^{(ij)}t_{a}^{E(ij)},
\label{tildezw}
\end{eqnarray}
with $t_{a}^{E(ij)}\tilde{z}_{a}^{(ij)}=0$. Further distinction is made in $\tilde{w}^{(ij)}$ between the contributions $\tilde{w}^{u(ij)}$  from the displacement of the centers and $
\tilde{w}^{\omega (ij)}$ from rotations, i.e.
\begin{eqnarray}
\tilde{w}^{u(ij)} &=&\left[ \tilde{u}_{l}^{(i)}-\tilde{u}_{l}^{(j)}
\right] t_{l}^{E(ij)},  \notag  \label{tildewu} \\
\tilde{w}^{\omega (ij)} &= &\frac{D}{2}\epsilon _{lbc}\left[ \tilde{
\omega}_{b}^{(i)}+\tilde{\omega}_{b}^{(j)}\right] n_{c}^{(ij)}t_{l}^{E(ij)},\notag\\
\tilde{w}^{(ij)} &=&\tilde{w}^{u(ij)}+\tilde{w}^{\omega (ij)}.
\label{tildewom}
\end{eqnarray}
We organize the contacts in sets depending on the polar angle
$\theta$ with respect to the major principal stress, as axial symmetry in the applied stress and isotropy in
the contact orientation make the displacement fluctuations
depend on it. We denote by $S_{\overline{\theta}}$ the set of polar angle $\overline{\theta}$, and consider \emph{equally
  oriented} all contacts that belong to it.

  \begin{figure}[t]
{\small 
}
\par
\begin{center}
{\small \includegraphics[height=10cm]{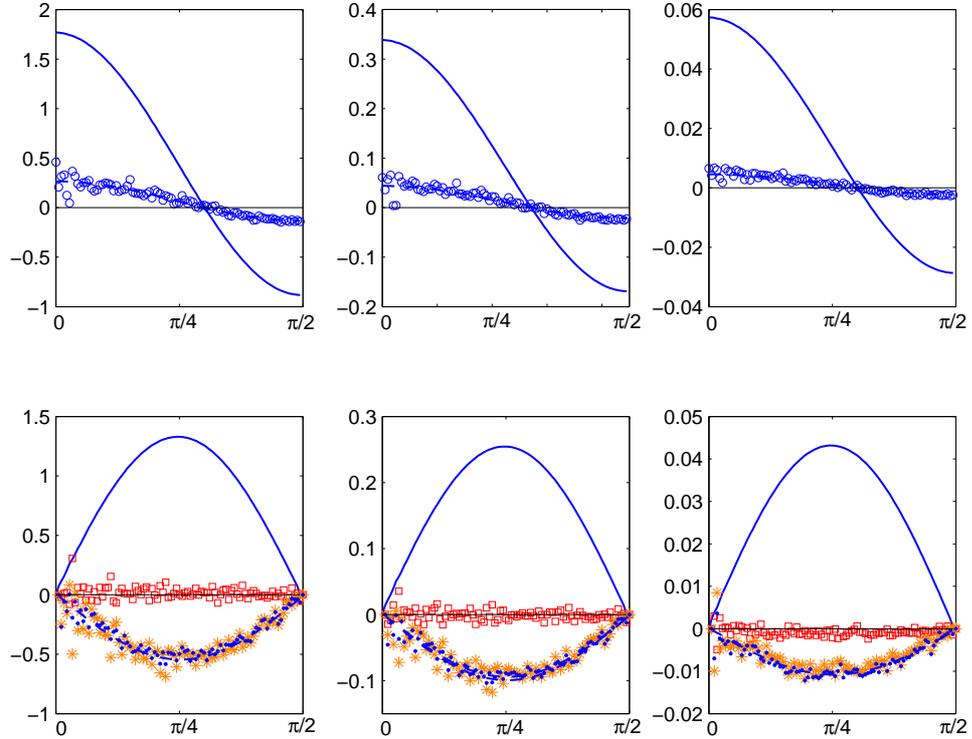}  }
\end{center}
\caption{{\protect\footnotesize {Normal (first row) and tangential
(second row) contact displacements for sample A, averaged over equally
oriented contacts, at $p=10KPa$ (left), $p=10^3
      KPa$ (center) and $p=10^5 KPa$. The results are in diameter units,
      multiplied by $10^{4}$. Solid lines indicate the average strain
contribution; symbols: $\bullet$= $<\tilde{w}>$; *= $<\tilde{w}^{\protect
\omega}>$; $\Box=<\tilde{w}^{u}>$.}}}
\label{FigdisplA}
\end{figure}
\begin{figure}[t]
{\small 
}
\par
\begin{center}
{\small \includegraphics[height=10cm]{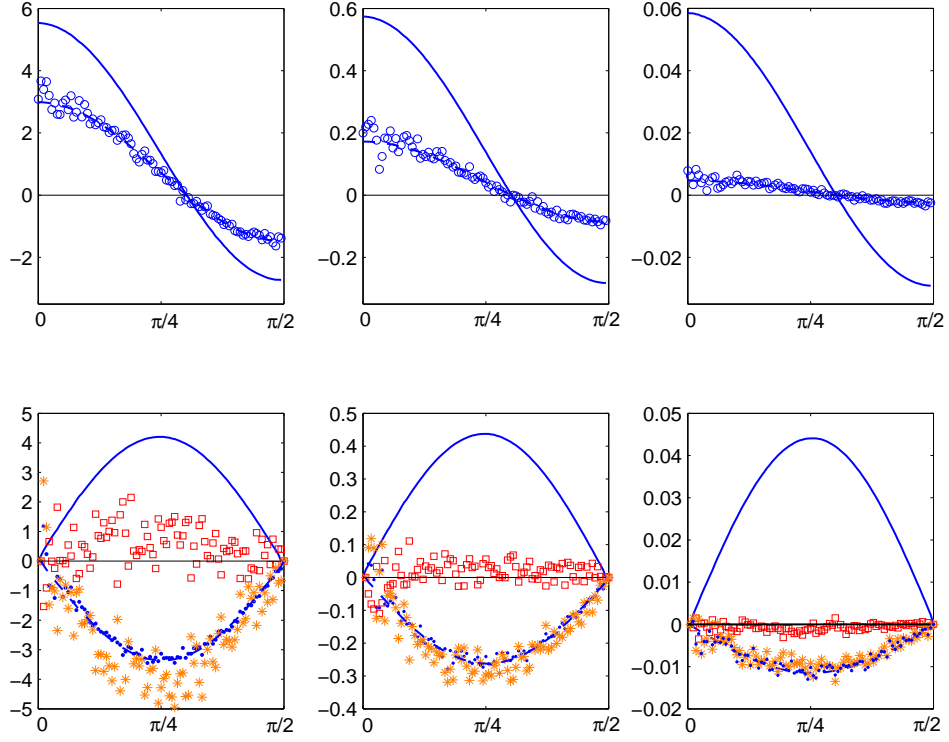}  }
\end{center}
\caption{{\protect\footnotesize {Normal (first row) and tangential
(second row) contact displacements for sample C. Same symbols as in Figure \ref{FigdisplA}.}}}
\label{FigdisplB}
\end{figure}

  Firstly, in Figures \ref{FigdisplA} and
\ref{FigdisplB} we plot the fluctuations averaged over each set, for sample $A$ and
$C$, respectively, with the polar angle made vary between $0$ and $90$
 degrees. Sample $A$ is also representative of the behavior of sample
$B$, which has a similar coordination number; in the same way, sample $C$ is also
representative of sample $D$. That is, the
displacement fluctuations chiefly depend on coordination number. Set by set, the
$\tilde{z_{a}}$'s are zero on average. Open and filled dots
represent, in turn, the average of $\tilde{h}^{(ij)}$ and $\tilde{w}^{(ij)}$, which
we denote by $\langle \tilde{h}\rangle _{\theta }$ and $\langle \tilde{
w}\rangle _{\theta }$. These are found to be proportional to the
displacement induced by the average strain and aligned with them. As a result, it is possible to write that
\begin{eqnarray}
\langle \tilde{h}\rangle _{\theta } &=&\alpha _{N}h^{E(ij)},  \notag \\
\langle \widetilde{w}\rangle _{\theta } &=&\alpha
_{T}s_{a}^{E(ij)}t_{a}^{E(ij)},  
\label{AvFl}
\end{eqnarray}
with $\alpha _{N}$ and $\alpha _{T}$ the proportionality
factors. $\alpha _{N}$ and $\alpha _{T}$ are negative, because the fluctuations counteract the average
strain. Their value is reported in Table 1, which shows that the tangential fluctuations are significantly more
effective than the radial ones in relaxing the system. Contact by contact, $\tilde{h}^{(ij)}$ and $\tilde{w}^{(ij)}$
may significantly differ from their average $\langle \tilde{h}
\rangle _{\theta }$ and $\langle \widetilde{w}\rangle
_{\theta }$. If the
difference from the average is called ``residual'' and labelled by a superscript R, then
\begin{eqnarray}
\tilde{h}^{(ij)} &=&\alpha _{N}h^{E(ij)}+h^{R(ij)},  \notag \\
\widetilde{w}^{(ij)} &=&\alpha _{T}s_{a}^{E(ij)}t_{a}^{E(ij)}+w_{a}^{R(ij)}t_{a}^{E(ij)}.
\label{residual2}
\end{eqnarray} 
As $\alpha _{N}$ and $\alpha _{T}$ are independent of the polar angle, $\langle \tilde{h}
\rangle _{\theta }$ and $\langle \widetilde{w}\rangle
_{\theta }$ are determined by the average structure of the
assembly.

\normalsize 

Secondly, we want to understand how the displacement fluctuations determine the
macroscopic behavior of the assemblies. As they depend on the contact
overlap, both the normal and the tangential contact stiffness fluctuate
about an average, and one can write that
\begin{eqnarray}
K_{N}^{(ij)} &=&<K_{N}>+\tilde{K}_{N}^{(ij)},  \notag \\
K_{T}^{(ij)} &=&<K_{T}>+\tilde{K}_{T}^{(ij)}.  \label{Stiff}
\end{eqnarray}
\begin{table}[h]
{\normalsize \newpage }{\tiny 
\begin{tabular}{|lccc|lccc|}
\hline
Sample A &  &  &                                &Sample B &  & &  \\ 
p[KPa]& $\alpha _{N}$ & $\alpha _{T}$ & $
\alpha_{T}^{\omega}$& p[KPa] & $\alpha _{N}$ & $\alpha _{T}$ & $ \alpha_{T}^{\omega}$ \\ \hline
$10$              & -0.16 & -0.41 & -0.46& $10$              & -0.16 & -0.46 & -0.48\\
$10\sqrt{10}$     & -0.14 & -0.42 & -0.43& $10\sqrt{10}$     & -0.16 & -0.45 & -0.47\\ 
$10^{2}$          & -0.14 & -0.39 & -0.40& $10^{2}$          & -0.15 & -0.44 & -0.46\\ 
$10^{2}\sqrt{10}$ & -0.14 & -0.4  & -0.40& $10^{2}\sqrt{10}$ & -0.15 & -0.43 & -0.44\\ 
$10^{3}$          & -0.13 & -0.37 & -0.38& $10^{3}$          & -0.15 & -0.41 & -0.43\\ 
$10^{3}\sqrt{10}$ & -0.13 & -0.36 & -0.36& $10^{3}\sqrt{10}$ & -0.14 & -0.39 & -0.41\\ 
$10^{4}$          & -0.10 & -0.33 & -0.33& $10^{4}$          & -0.13 & -0.36 & -0.36\\ 
$10^{4}\sqrt{10}$ & -0.09 & -0.30 & -0.23& $10^{4}\sqrt{10}$ & -0.10 & -0.33  & -0.33\\ 
$10^{5}$          & -0.07 & -0.25 & -0.31& $10^{5}$          & -0.08 & -0.28 & -0.26\\ \hline
Sample C &  &  &                                &Sample D &  & &   \\ 
$10$              & -0.54 & -0.80 & -1.04& $10$              & -0.49 & -0.78 & -0.96\\ 
$10\sqrt{10}$     & -0.48 & -0.76 & -0.96& $10\sqrt{10}$     & -0.48 & -0.77 & -0.91\\ 
$10^{2}$          & -0.44 & -0.73 & -0.90& $10^{2}$          & -0.45 & -0.74 & -0.89\\ 
$10^{2}\sqrt{10}$ & -0.36 & -0.68 & -0.81& $10^{2}\sqrt{10}$ & -0.40 & -0.71 & -0.82\\ 
$10^{3}$          & -0.30 & -0.60 & -0.71& $10^{3}$          & -0.34 & -0.67 & -0.77\\ 
$10^{3}\sqrt{10}$ & -0.24 & -0.53 & -0.60& $10^{3}\sqrt{10}$ & -0.29 & -0.61 & -0.69\\
$10^{4}$          & -0.18 & -0.45 & -0.49& $10^{4}$          & -0.24 & -0.54  & -0.59\\
$10^{4}\sqrt{10}$ & -0.13 & -0.35 & -0.37& $10^{4}\sqrt{10}$ & -0.19 & -0.46 & -0.48\\ 
$10^{5}$          & -0.09 & -0.26 & -0.27& $10^{5}$          & -0.15 & -0.37 & -0.37\\ \hline
\end{tabular}
}
\caption{{\protect\footnotesize {Contribution from the structural fluctuations to the
stress for samples A, B, C, D.}}}
\end{table}
Once (\ref{DForce}), (\ref{displ}), (\ref{dtilde}
), (\ref{tildezw}), (\ref{residual2}) and (\ref{Stiff}) have been inserted in
(\ref{defStress}), and sums over grains have been transformed into sums over contacts spanning the
sets $S(\theta)$, one obtains 
\begin{eqnarray*}
\Delta \sigma _{rm}\frac{V}{D} 
&=&\sum_{\theta}\sum_{(ij) \in S_{\theta}}\left[
   <K_{N}>+\tilde{K}_{N}^{(ij)}\right] \left[
   \left(1+\alpha_{N}\right)h^{E(ij)}+h^{R(ij)}\right]n_{r}^{(ij)}n_{m}^{(ij)}
  \notag \\
&+&\sum_{\theta}\sum_{(ij)\in S_{\theta}}\left[
<K_{T}>+\tilde{K}_{T}^{(ij)}\right] \left[\delta_{rl}-
n_{r}^{(ij)}n_{l}^{(ij)}\right]
\notag\\
&&\;\;\;\;\;\;\;\;\;\;\;\times \left[ \left(1+\alpha_{T}\right)s_{l}^{E(ij)}+w^{R(ij)}t_{l}^{E(ij)}+\tilde{z}
_{l}^{(ij)}\right] n_{m}^{(ij)}.  
\label{Stress2}
\end{eqnarray*}

{\small 
\begin{figure}[t]
{\small 
}
\par
\begin{center}
{\small \includegraphics[height=10cm]{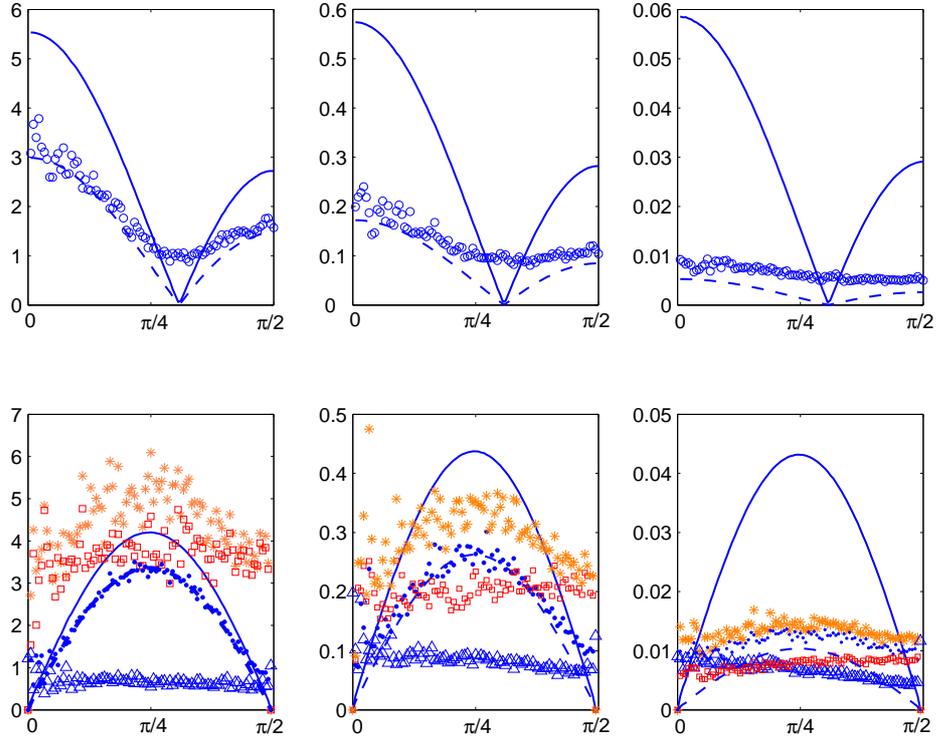}  }
\end{center}
\caption{{\protect\footnotesize {Modulus of normal (first row) and
tangential (second row) contact displacements averaged over equally oriented
contacts and multiplied by $10^{4}$. Same symbols as in Figure \ref{FigdisplA}.}%
}}
\label{ModdisplB}
\end{figure}}

\begin{figure}[t]
{\small 
}
\par
\begin{center}
{\small \includegraphics[width=14cm]{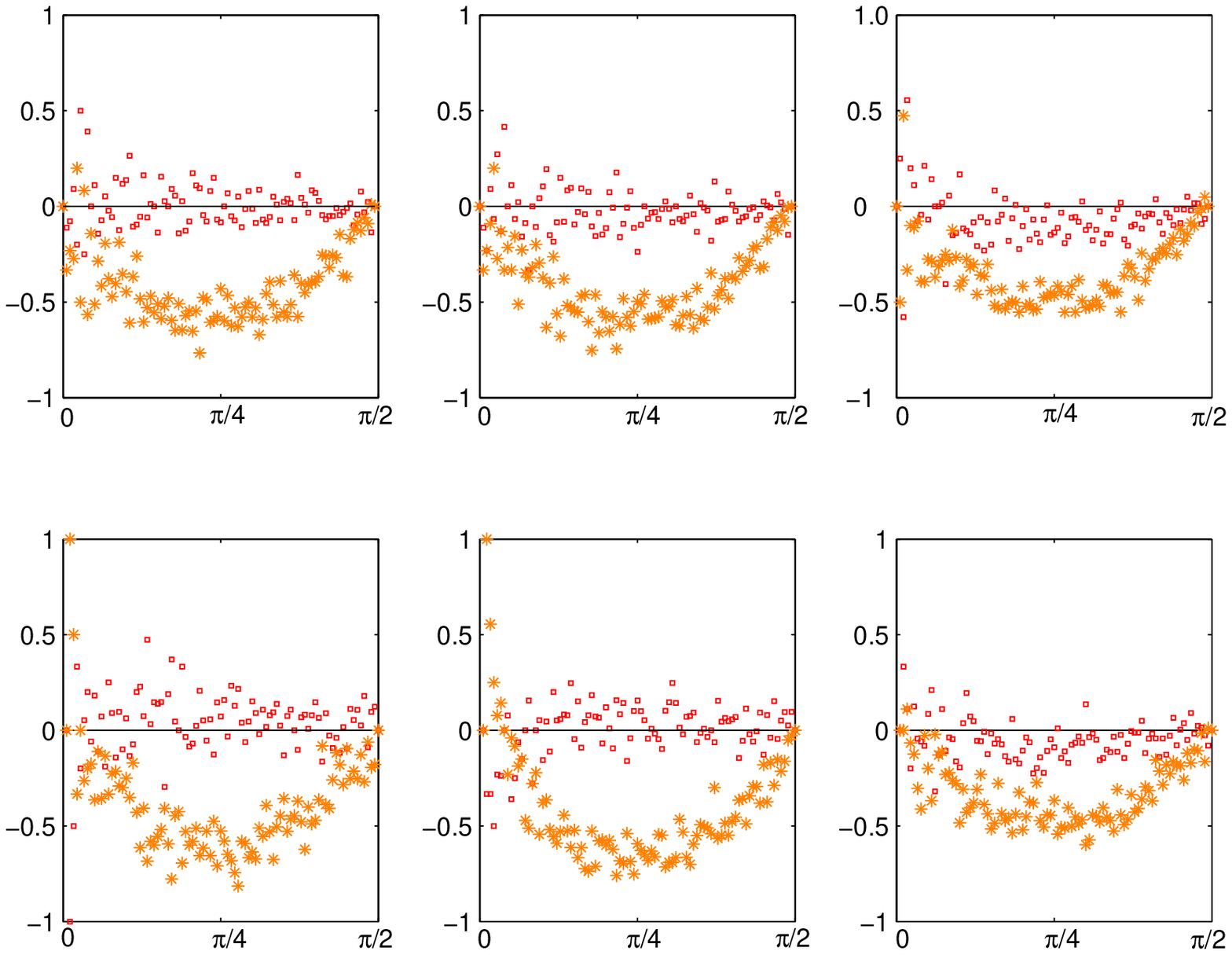}  }
\end{center}
\caption{{\protect\footnotesize {Sign of $\tilde{w}^{u}$($\Box$) and ($%
\tilde{w}^{\protect\omega}$)(*) over equally oriented contacts for sample
A(first row) and C (second row), at pressure $p=10$(left), $10^{3}$(center)
and $10^{5}KPa$(right). A positive (negative) value means the dominance of
the positive (negative) sign; the numerical value gives the fraction of
contacts that sign prevails by.}}}
\label{SignuOm}
\end{figure}
With respect to (\ref{defStress}), the factor $1/2$ has disappeared, as double
counting does not occur when the indexes are made vary over the contacts. Due to the definition of fluctuations, products of fluctuations in the contact stiffness, the residuals or the
$\tilde{z}_{l}$'s times the average stiffness, the
$h^{E(ij)}$ or the $s_{l}^{E(ij)}$'s give null
terms. In what remains, products of
fluctuations give contributions whose ratio to the total stress ranges between $10^{-3}$
  and $10^{-7}$, and which can thus be neglected. As a consequence,
\begin{eqnarray}
\Delta \sigma _{lm} \frac{V}{D}&\simeq &<K_{N}>\left( 1+\alpha _{N}\right)\sum_{\theta}\sum_{(ij) \in S_{\theta}}
h^{E(ij)}n_{l}^{(ij)}n_{m}^{(ij)}  \notag \\
&+&<K_{T}>\left( 1+\alpha
_{T}\right)\sum_{\theta}\sum_{(ij) \in S_{\theta}} s_{l}^{E(ij)}n_{m}^{(ij)}. 
\label{scompStress2}
\end{eqnarray}
Expression (\ref{scompStress2}) makes it explicit that the average structure is defined by the average contact stiffness and the average contact distribution. Local deviations from the average structure give rise to the residuals and the $\tilde{z}$'s, which disappear at the macroscopic scale. However, their numerical value is comparable to that of the average fluctuations, as Figure \ref{ModdisplB} proves. In this figure, the absolute value of the residuals, averaged over equally oriented contacts, is inferred by comparing the absolute value of normal and tangential fluctuations, averaged over the contact sets, with the absolute value of the average. All quantities decrease with increasing coordination number, as the forces induced by the average strain balance more and more easily, but the residuals, especially in the tangent plane, and the $\tilde{z}$'s decrease more slowly than the average fluctuations. This means that the issue of local equilibrium is less sensitive than that of the average structure to coordination number.

As in (\ref{ContStress}), (\ref{scompStress2}) can be transformed into an
integral over the solid angle. As a result,
\begin{equation}
G=\frac{1}{10}\left[6 \left( 1+\alpha _{N}\right) +9  \left( 1+\alpha
_{T}\right)\frac{<K_{T}>}{<K_{N}>}\right]B^{E},
\label{exprG}
\end{equation}
which states that the prediction of the shear modulus requires that of $\alpha _{N}$
and $\alpha _{T}$. 

Tangential fluctuations are contributed by both center displacements and rotations. The average over the contact sets of the $\widetilde{w}^{\omega (ij)}$ and $\widetilde{w}^{u(ij)}$, introduced in (\ref{tildewom}), is
also determined, and reported in Figures \ref{FigdisplA} and
\ref{FigdisplB}. This time, significant deviations
from the proportionality to $s_{a}^{E(ij)}t_{a}^{E(ij)}$ are observed,
and especially the $\widetilde{w}^{u(ij)}$'s distribute rather isotropically. In Agnolin and Kruyt (2006), a strict
proportionality is still observed in two-dimensional random
assemblies of polydisperse disks with identical contact stiffness. Therefore, the poor correlation
found for spheres is the specific signature of inhomogeneity in the contact
stiffness. A closer insight into the role of center displacements and rotations is
given by averaging $sign\left( \tilde{w}^{u}\right)$ and
$sign\left(\tilde{w}^{\omega }\right)$ over the sets $S(\theta)$. As a definition,
\begin {eqnarray*}
\langle sign\left( \tilde{w}^{u}\right)\rangle _{\theta}&=&\sum_{ij\in
  S(\theta)}\frac{sign\left(
  \tilde{w}^{u}\right)}{N(S_{\theta})}=\frac{N^{+}\left(
  \tilde{w}^{u}\right)-N^{-}\left( \tilde{w}^{u}\right)}{N(S_{\theta})},\\
\langle sign\left( \tilde{w}^{\omega}\right)\rangle _{\theta}&=&\sum_{ij\in
  S(\theta)}\frac{sign\left( \tilde{w}^{\omega}\right)}{N(S_{\theta})}=\frac{N^{+}\left( \tilde{w}^{\omega}\right)-N^{-}\left( \tilde{w}^{\omega}\right)}{N(S_{\theta})},
\end{eqnarray*}
where $N(S_{\theta})$ is the number of contacts in the set $S_{\theta}$, and
$N^{+}(\,)$ and $N^{-}(\,)$ give the number of contacts at which the bracketed
quantity is,
in turn, positive and negative. A (positive) negative result means that at the chosen
polar angle the fluctuations mainly (do not) counteract the displacements induced by the
average strain, while the numerical value specifies the percent of contacts the dominant sign
prevails by. Figure \ref{SignuOm}, which refers to sample $C$, shows
that rotations mainly oppose the average strain, the more efficiently the larger the displacements induced by the average strain. Exceptions
are the closest orientations to the main compression. However, the corresponding contact sets are not statistically representative, because of the low number of contacts they gather due to isotropy in the distribution of
the contact orientation. As far as the center displacements are concerned, these do not counteract the average strain, especially at small coordination numbers. From (\ref{scompStress2}), one can also say that
\begin{eqnarray}
\alpha _{N} &=&\Delta \sigma _{lm}(\widetilde{h})/\Delta \sigma _{lm}(h^{E}),
\notag \\
\alpha _{T} &=&\Delta \sigma _{lm}(\tilde{w})/\Delta \sigma _{lm}(s^{E}),
\label{spec}
\end{eqnarray}
where brackets include the kinematic term whose contribution to the stress is
considered. By analogy with (\ref{spec}), we introduce the numerical
factors $\alpha _{T}^{\omega }$ and $\alpha _{T}^{u}$, namely
\begin{eqnarray}
\alpha _{T}^{\omega} &\doteq &\Delta \sigma _{lm}(\widetilde{w}^{\omega })/\Delta \sigma _{lm}(s^{E}),  \notag \\
\alpha _{T}^{u} &\doteq &\Delta \sigma _{lm}(\widetilde{w}^{u})/\Delta \sigma _{lm}(s^{E}),
\label{atUatOm}
\end{eqnarray}
which allow to measure the contribution to
the macroscopic behavior from rotations and center displacements. $\alpha _{T}^{\omega}$ is also reported in Table 1, averaged
over the directions 11, 22 and 33, while $\alpha _{T}^{u}$ can be inferred, as
$\alpha _{T}^{u}=\alpha _{T}-\alpha _{T}^{\omega }$. The contribution from the center
displacements is much smaller than that due to rotations. Moreover, the
relaxation induced along the tangential direction comes entirely from rotations, with
the exception of sample $A$ at the pressure of $10^{4}\sqrt{10}KPa$ and sample
$B$ at the pressure of $10^{5}\sqrt{10}KPa$. 

\section{Predictions}\label{Predictions}
Reliable estimates of the shear modulus need to incorporate displacement
fluctuations compatible with (\ref{AvFl}). This is the case in the 1-fluctuating-particle
approach (1FP) and in the Pair-Fluctuation (PF) approach, both based on the balance of force and torque on a small-sized subassembly. In this section, the two methods are briefly recalled, numerically emplyed and compared.

The 1-fluctuating-particle
approach (1FP) is put forward in Kruyt and Rothenburg (2002) to numerically predict the elastic
moduli of two-dimensional assemblies of non-rotating frictional disks. In this approach, the balance of force and moment
is solved grain by grain in the assembly. In more detail, the chosen grain alone, say $i$, is allowed to fluctuate, while its neighbors are
compelled to move in accordance with the average strain. As a result, the relative displacement $u_{t}^{(ia)}$ differs from (\ref{displ}), as 
\begin{equation}
u_{t}^{(ia)}=DE_{tm}n_{m}^{(ia)}+\tilde{u}_{t}^{(i)}+\frac{D}{2}\epsilon
_{tbc}\tilde{\omega}_{b}^{(i)}n_{c}^{(ia)}, \label{displia}
\end{equation}
and the equilibrium equations for the chosen grain, i.e.
\begin{eqnarray}
&&\sum_{a=1}^{N(i)}\left[ \left( K_{N}^{(ia)}-K_{T}^{(ia)}\right)
n_{r}^{(ia)}n_{t}^{(ia)}+K_{T}^{(ia)}\delta _{rt}\right] u_{t}^{(ia)}=0, 
\notag  \label{eqFi} \\
&&\sum_{a=1}^{N(i)}\epsilon _{vgr}n_{g}^{(ia)}\left[ \left(
K_{N}^{(ia)}-K_{T}^{(ia)}\right) n_{r}^{(ia)}n_{t}^{(ia)}+K_{T}^{(ia)}\delta
_{rt}\right] u_{t}^{(ia)}=0,  \label{eqMi}
\end{eqnarray}
can be solved for its fluctuations if the average strain is considered
known. The displacement at the contact point between interacting grains, say $i$ and $j$, is conveniently expressed using the notation 
\begin{eqnarray}
\tilde{u}_{r}^{(i)}-\tilde{u}_{r}^{(j)} &=&\beta _{rs}^{(ij)}e_{s},  \notag\\
\tilde{\omega}_{r}^{(i)}+\tilde{\omega}_{r}^{(j)} &=&\gamma
_{rs}^{(ij)}e_{s}.\label{betaFluct}
\end{eqnarray}
In (\ref{betaFluct}), the vector $\mathbf{e}$ has
components 
\begin{equation*}
e_{1}=E_{11},\:e_{2}=E_{22},\:e_{3}=E_{33},
\end{equation*}
and $\beta _{rs}^{(ij)}$ and $\gamma _{rs}^{(ij)}$, which describe the stiffness and geometry of their neighborhood, are inferred from the components of the matrices associated with the two problem of equilibrium (\ref{eqMi}) for grain $i$ and $j$. The contact radial shortening $\tilde{h}^{(ij)}$ and
the tangential
relative displacement $\tilde{\mathbf{s}}^{(ij)}$ that correspond to
(\ref{betaFluct}) are
\begin{eqnarray}
\tilde{h}^{(ij)} &=&\beta _{rm}^{(ij)}e_{m}n_{r}^{(ij)},  \notag \\
\tilde{s}_{r}^{(ij)} &=&\beta _{sm}^{(ij)}e_{m}\left( \delta
_{sr}-n_{s}^{(ij)}n_{r}^{(ij)}\right) +\frac{D}{2}\epsilon _{rbc}\gamma
_{bm}^{(ij)}e_{m}n_{c}^{(ij)}.  \label{deltaC}
\end{eqnarray}
On the base of (\ref{relaz}) and (\ref{4Ctens}), the contributions $\widetilde{C}_{1111}$ and $\tilde{C}
_{1122} $ from the fluctuation of the pairs to the forth
order tensor and $\tilde{G}$ to the shear modulus are
\begin{eqnarray}
\tilde{C}_{1111}&=&\frac{D}{V} \sum_{(ij)=1}^{C}\left[
\left( K_{N}^{(ij)}-K_{T}^{(ij)}\right) \beta
_{r1}^{(ij)}n_{r}^{(ij)}n_{1}^{(ij)}\right. \notag\\
&&\left. \;\;\;\;+K_{T}^{(ij)}\left( \beta _{11}^{(ij)}+\frac{D}{2}
\epsilon _{1bc}\gamma _{b1}^{(ij)}n_{c}^{(ij)}\right) \right] n_{1}^{(ij)}, \notag\\
\tilde{C}_{1122}&=&\frac{D}{V} \sum_{(ij)=1}^{C}\left[
\left( K_{N}^{(ij)}-K_{T}^{(ij)}\right) \beta
_{r2}^{(ij)}n_{r}^{(ij)}n_{1}^{(ij)}\right.  \notag\\
&&\left. \;\;\;\;+K_{T}^{(ij)}\left( \beta _{12}^{(ij)}+\frac{D}{2}
\epsilon _{1bc}\gamma _{b2}^{(ij)}n_{c}^{(ij)}\right) \right]
n_{1}^{(ij)}, \notag\\
\tilde{G}&=&( \tilde{C}_{1111}-\tilde{C}_{1122}) /2,
\label{grosso}
\end{eqnarray}
where the sums are carried out over the contacts in the assembly.

\begin{figure}[h]
{\small 
}
\par
\begin{center}
{\small 
\includegraphics[width=7cm]{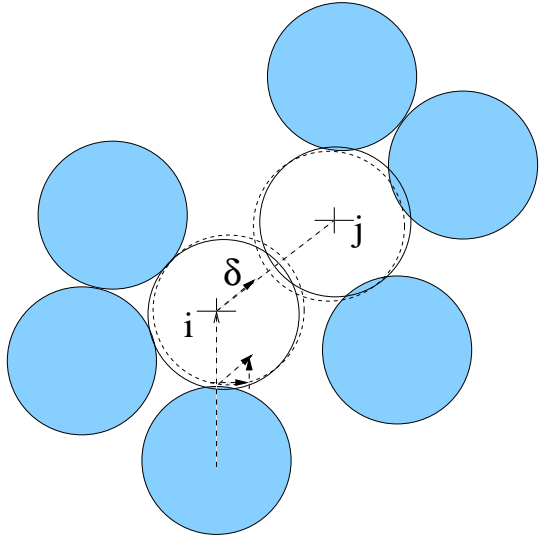}  }
\end{center}
\caption{{A pair of contacting grains $ij$ with their neighborhood in the PF approach.
The pair alone is allowed to fluctuate, while the grains in contact with
them are compelled to move in accordance with the average strain. A radial
relative displacement $\delta$ of the pair induces forces on the neighbors
that have non-zero normal and tangential component.}}
\label{Neigh}
\end{figure}

The pair fluctuation model (PF) is employed in Agnolin et al. (2005a)
and Jenkins et al. (2005) to
analytically predict the shear modulus of, in turn, two-dimensional systems
of frictional and frictionless disks and of three-dimensional assemblies of
frictionless spheres. In this method, the balance of force and torque is
solved pair by pair of contacting grains in the assembly. The pair alone is allowed to fluctuate, while its
neighborhood is compelled to move in accordance with the average strain, as shown
in Figure \ref{Neigh}. If $i$ and $j$ are the indexes of
the grains of the chosen pair, their relative displacement $u_{t}^{(ij)}$
is still given by (\ref{displ}), while (\ref{displia}) holds whenever a
neighbor of $i$ different from $j$ is considered, and
\begin{equation}
u_{t}^{(jb)}=DE_{tm}n_{m}^{(jb)}+\tilde{u}_{t}^{(j)}+\frac{D}{2}\epsilon
_{tbc}\tilde{\omega}_{b}^{(j)}n_{c}^{(jb)}  \label{displjb}
\end{equation}
is employed when $b\neq i$. With the average strain assigned, the system consisting of the equilibrium
equations (\ref{eqMi}) for grains $i$ and 
\begin{eqnarray}
&&\sum_{b=1}^{N(j)}\left[ \left( K_{N}^{(jb)}-K_{T}^{(jb)}\right)
n_{r}^{(jb)}n_{t}^{(jb)}+K_{T}^{(jb)}\delta _{rt}\right] u_{t}^{(jb)}=0, 
\notag  \label{eqFi} \\
&&\sum_{b=1}^{N(j)}\epsilon _{vgr}n_{g}^{(jb)}\left[ \left(
K_{N}^{(jb)}-K_{T}^{(jb)}\right) n_{r}^{(jb)}n_{t}^{(jb)}+K_{T}^{(jb)}\delta
_{rt}\right] u_{t}^{(jb)}=0  \label{eqMj}
\end{eqnarray}
for grain $j$ can be solved for the fluctuations of the pair if the average strain is
considered known. Expressions (\ref{deltaC}) and (\ref{grosso}) can still be employed to express the relative displacement and the contribution from the pair to the elastic moduli, but this time $\beta _{rs}^{(ij)}$ and $\gamma _{rs}^{(ij)}$ are inferred from the components of the solving matrix associated with the problem of equilibrium of the pair. It is finally noticed that the fluctuations of the individual particles are not unique in the PF method, as each
force-carrying grain belongs to more than one pair. 

The relative displacements induced by the estimated fluctuations are averaged over equally oriented contacts. Due to the small size of the representative domains, such averages are not nicely proportional anymore to the relative displacements induced by the average strain, especially in the tangential plane. The values of $\alpha_{N}$ and $\alpha_{T}$, which measure the contribution from the fluctuations to the macroscopic stress, are thus determined using (\ref{spec}), and plotted in Figure \ref{alphaPQ} for samples $A$ and $C$. 
The corresponding 1FP and PF estimates
of the shear modulus are plotted in Figure \ref{Qual} in
terms of their ratio to the effective value. The results bound the performance that could be achieved by the corresponding analytical solutions, and emphasize a
remarkable improvement with respect to the average strain assumption, especially
for sample A. In the case of sample
C, in the worst case the overestimate is reduced from 3.3 to about 1.5: further improvement
would require to incorporate, at least, the fluctuations of the first
neighbors. The results allow to conclude that the fluctuations which most affect the macroscopic
behavior do correlate within short
distance, in accordance with Gaspar and Koenders (2001) and Agnolin and
Kruyt (2006), the more the larger the coordination number. In contrast with the results obtained by Agnolin and
Kruyt (2006) in two dimensions, the tangential fluctuations are better predicted than the radial
ones. 

\subsection{1FP vs PF}\label{vs}
The PF approach gives worse estimates than the 1FP method, even
though it relaxes more degrees of freedom. In order to get a better insight into this finding, we focus on two contacting grains, say $i$ and $j$, and compare the equilibrium equations that result for the pair
from the two approaches. Three quantities describe the stiffness of
the neighborhood of grain $i$ in (\ref{eqMi}), namely 
\begin{eqnarray}
I_{rt}^{(ij)} &\doteq&\sum_{a=1}^{N(i)}K_{rt}^{(ia)},  \notag \\
J_{rts}^{(ij)} &\doteq&\sum_{a=1}^{N(i)}n_{r}^{(ia)}K_{ts}^{(ia)},  \notag \\
P_{rtsg}^{(ij)} &\doteq&\sum_{a=1}^{N(i)}n_{r}^{(ia)}K_{ts}^{(ia)}n_{g}^{(ia)},
\label{structuralSums}
\end{eqnarray}
which we have labelled by $(ij)$ because of our focus on the pair. Analogous quantities, say $I_{rt}^{(ji)}$, $J_{rts}^{(ji)}$ and $P_{rtsg}^{(ji)}$,
stem from the equilibrium of grain $j$. Sums like these, whose largest order
depends on the size of the subsystem on which equilibrium is enforced, were first studied by Koenders (1987), who called them ``structural sums''. If the equations
of balance of the force and torque are, in turn, subtracted from and summed
to each other, one
obtains in the 1FP case:
\begin{eqnarray}
&&2\left\{\;-\left[ I_{rt}^{(ij)}+I_{rt}^{(ji)}\right]\left(
  \tilde{u}_{t}^{(i)}-\tilde{u}_{t}^{(j)}\right)\,- \,\left[ I_{rt}^{(ij)}-I_{rt}^{(ji)}\right]\left( \tilde{u}_{t}^{(i)}+\tilde{u}_{t}^{(j)}\right)\;\right\}\notag\\
&+&\epsilon _{sbc}D\left\{\;
 \left[ J_{brs}^{(ij)}-J_{brs}^{(ji)}\right] \left(\tilde{\omega}
    _{c}^{(i)}+\tilde{\omega} _{c}^{(j)}\right)\,+\,\epsilon _{sbc}D\left[ J_{brs}^{(ij)}+J_{brs}^{(ji)}\right] \left( \tilde{\omega}
    _{c}^{(i)}-\tilde{\omega} _{c}^{(j)}\right)\;\right\}  \notag\\
&=&-4D\left[ J_{crs}^{(ij)}-J_{crs}^{(ji)}\right]E_{sc},\notag\\
&&2\epsilon _{sbc}\left\{\;-\left[ J_{bct}^{(ij)}-J_{bct}^{(ji)}\right]\left(
  \tilde{u}_{t}^{(i)}-\tilde{u}_{t}^{(j)}\right) \,-\,\left[ J_{bct}^{(ij)}+J_{bct}^{(ji)}\right]\left(
  \tilde{u}_{t}^{(i)}+\tilde{u}_{t}^{(j)}\right)\;\right\}\notag\\
&+&\epsilon _{sbv}\epsilon _{drc}D\left\{\;\left[ P_{bvdr}^{(ij)}+P_{bvdr}^{(ji)}\right] \left(\tilde{\omega}
    _{c}^{(i)}+\tilde{\omega} _{c}^{(j)}\right) \,+\,\left[
  P_{bvdr}^{(ij)}-P_{bvdr}^{(ji)}\right] \left( \tilde{\omega}
    _{c}^{(i)}-\tilde{\omega} _{c}^{(j)}\right)\;\right\} \notag\\ 
&=&-\epsilon _{sbc}4D\left[ P_{bcst}^{(ij)}+P_{bcst}^{(ji)}\right]E_{st}.
\label{solvSys1}
\end{eqnarray}
In the PF case, the resulting equations differ from (\ref{solvSys1}) in a few terms, which are the only ones written:
\begin{eqnarray}
&&2\left\{ \; -\left[
    I_{rt}^{(ij)}+I_{rt}^{(ji)}+2K_{rt}^{(ji)}\right]\left(\tilde{u}_{t}^{(i)}-\tilde{u}_{t}^{(j)}\right)\,-...\;\right\}\notag\\
&+&\epsilon _{sbc}D\left\{\;\left[J_{brs}^{(ij)}-J_{brs}^{(ji)}+2n^{(ij)}_{b}K^{(ij)}_{rs}\right] 
\left(\tilde{\omega}_{c}^{(i)}+\tilde{\omega} _{c}^{(j)}\right)\,-...\;\right\}\notag\\
&=&...,\notag\\
&&2\epsilon _{sbc}\left\{\;-\left[ J_{bct}^{(ij)}-J_{bct}^{(ji)}+2n^{(ij)}_{b}K^{(ij)}_{ct}\right]\left(
  \tilde{u}_{t}^{(i)}-\tilde{u}_{t}^{(j)}\right)- \,...\;\right\} \notag\\
&+&\epsilon _{sbv}\epsilon _{drc}D\left\{\;\left[ P_{bvdr}^{(ij)}+P_{bvdr}^{(ji)}+2n^{(ij)}_{b}K^{(ij)}_{vd}n^{(ij)}_{r}\right] \left(\tilde{\omega}
    _{c}^{(i)}+\tilde{\omega} _{c}^{(j)}\right)- \,...\;\right\}\notag\\
&=&....
\label{solvSys2}
\end{eqnarray}
On the other hand, if the equations
of force balance and those
of torque balance are, in turn, summed to and subtracted
from each other, one obtains in the 1FP case:
\begin{eqnarray}
&&2\left\{ \;-\left[ I_{rt}^{(ij)}-I_{rt}^{(ji)}\right]\left(
  \tilde{u}_{t}^{(i)}-\tilde{u}_{t}^{(j)}\right)\,- \,\left[ I_{rt}^{(ij)}+I_{rt}^{(ji)}\right]\left( \tilde{u}_{t}^{(i)}+\tilde{u}_{t}^{(j)}\right)\;\right\}\notag\\
&+&\epsilon _{sbc}D\left\{\;
 \left[ J_{brs}^{(ij)}+J_{brs}^{(ji)}\right] \left(\tilde{\omega}
    _{c}^{(i)}+\tilde{\omega} _{c}^{(j)}\right)\,+\,\epsilon _{sbc}D\left[ J_{brs}^{(ij)}-J_{brs}^{(ji)}\right] \left( \tilde{\omega}
    _{c}^{(i)}-\tilde{\omega} _{c}^{(j)}\right)\;\right\}  \notag\\
&=&-4D\left[ J_{crs}^{(ij)}+J_{crs}^{(ji)}\right]E_{sc},\notag\\
&&2\epsilon _{sbc}\left\{\;-\left[ J_{bct}^{(ij)}+J_{bct}^{(ji)}\right]\left(
  \tilde{u}_{t}^{(i)}-\tilde{u}_{t}^{(j)}\right) \,-\,\left[ J_{bct}^{(ij)}-J_{bct}^{(ji)}\right]\left(
  \tilde{u}_{t}^{(i)}+\tilde{u}_{t}^{(j)}\right)\;\right\}\notag\\
&+&\epsilon _{sbv}\epsilon _{drc}D\left\{\;\left[ P_{bvdr}^{(ij)}-P_{bvdr}^{(ji)}\right] \left(\tilde{\omega}
    _{c}^{(i)}+\tilde{\omega} _{c}^{(j)}\right) \,+\,\left[
  P_{bvdr}^{(ij)}+P_{bvdr}^{(ji)}\right] \left( \tilde{\omega}
    _{c}^{(i)}-\tilde{\omega} _{c}^{(j)}\right)\;\right\} \notag\\ 
&=&-\epsilon _{sbc}4D\left[ P_{bcst}^{(ij)}-P_{bcst}^{(ji)}\right]E_{st},
\label{solvSys3}
\end{eqnarray}
while in the PF case:
\begin{eqnarray}
&&2\left\{...\,- \,\left[ I_{rt}^{(ij)}+I_{rt}^{(ji)}-2K_{rt}^{(ij)}\right]\left( \tilde{u}_{t}^{(i)}+\tilde{u}_{t}^{(j)}\right)\;\right\}\notag\\
&+&\epsilon _{sbc}D\left\{\;
 ...\,+\,\epsilon _{sbc}D\left[ J_{brs}^{(ij)}-J_{brs}^{(ji)}-2n_{b}^{(ij)}K_{rs}^{(ij)}\right] \left(\tilde{ \omega}
    _{c}^{(i)}-\tilde{\omega} _{c}^{(j)}\right)\;\right\}  \notag\\
&=&\,...,\notag\\
&&2\epsilon _{sbc}\left\{\;... \,-\,\left[ J_{bct}^{(ij)}-J_{bct}^{(ji)}-2n_{b}^{(ij)}K_{ct}^{(ij)}\right]\left(
  \tilde{u}_{t}^{(i)}+\tilde{u}_{t}^{(j)}\right)\;\right\}\notag\\
&+&\epsilon _{sbv}\epsilon _{drc}D\left\{\;...\,+\,\left[
  P_{bvdr}^{(ij)}+P_{bvdr}^{(ji)}-2n_{b}^{(ij)}K_{vd}^{(ij)}n_{r}^{(ij)}\right] \left( \tilde{\omega}
    _{c}^{(i)}-\tilde{\omega} _{c}^{(j)}\right)\;\right\} \notag\\ 
&=&....
\label{solvSys4}
\end{eqnarray}
Comparison between equations (\ref{solvSys1}) and (\ref{solvSys2}) on one side, and
(\ref{solvSys3}) and (\ref{solvSys4}) on the other shows that the 1FP and the PF approach differ only in the
isolated contribution from the chosen contact $ij$. Our calculations prove that this contribution has a stiffening effect on the behavior of the equivalent continuum. 

{\small 
\begin{figure}[h]
{\small 
}
\par
\begin{center}
{\small 
\includegraphics[width=12cm]{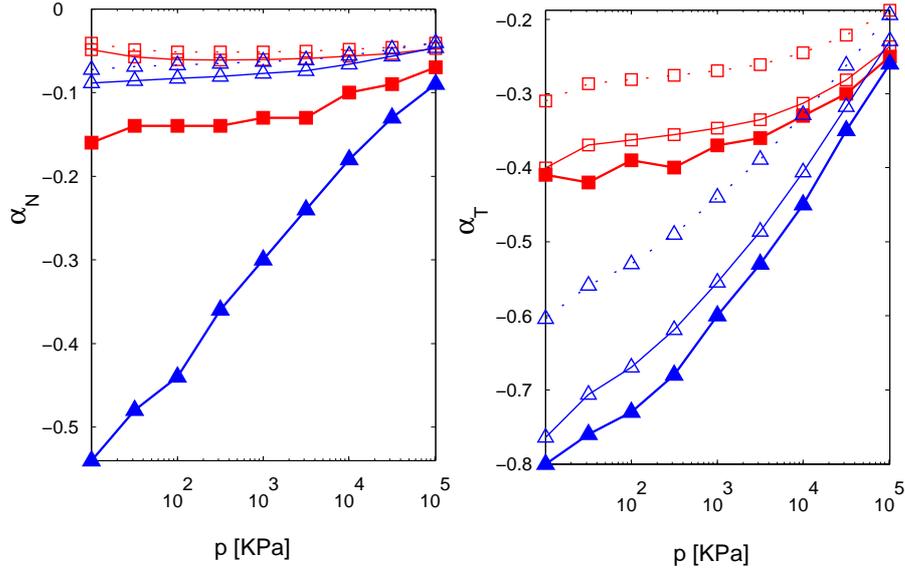}  }
\end{center}
\caption{{Estimated and measured value of $\protect\alpha_{N}$ and $\protect%
\alpha_{T}$ for sample A($\Box$) and C($\triangle$). Filled symbols: measured
values; empty symbols and dotted line: PF estimate; empty symbols and solid
line: 1FP estimate. }}
\label{alphaPQ}
\end{figure}
}

Additional information about the way this contribution enters the macroscopic behavior in the 1FP and FP model can be inferred once an averaging process has been applied on the solving equations. To this aim, use is made of the main findings in section \ref{SectionFluct}, where it has been shown that only the average of the fluctuations over equally oriented contacts is relevant to the macroscopic behavior; that this average is determined by the average
structure of the assembly, and, finally, that such an average structure depends on the average contact stiffness and the average geometry. On this base, it is feasible to substitute the local structural sums in (\ref{solvSys1}), (\ref{solvSys2}), (\ref{solvSys3}) and (\ref{solvSys4}) with their average over contacts with the same solid angle. As a result, the equations will give the average displacement over equally oriented contacts. We find it convenient to introduce the notation:
\begin{eqnarray}
\overline{I}_{rt}^{(ij)}&=&\langle K_{N}\rangle \text{
}n_{r}^{(ij)}n_{t}^{(ij)}+\langle K_{T}\rangle \text{ }\left( \delta
_{rt}-n_{r}^{(ij)}n_{t}^{(ij)}\right) \notag\\
&+&\langle K_{N}\rangle \text{ }\langle
\sum_{a=1,a\neq j}^{N(i)}n_{r}^{(ia)}n_{t}^{(ia)}\rangle +\langle K_{T}\rangle \text{
}\langle \sum_{a=1,a\neq j}^{N(i)}\left( \delta
_{rt}-n_{r}^{(ia)}n_{t}^{(ia)}\right) \rangle,\notag\\
\overline{J}_{rts}^{(ij)}&=&\langle K_{N}\rangle \text{ }n_{r}^{(ij)}n_{t}^{(ij)}n_{s}^{(ij)}
+\langle K_{T}\rangle \text{ }n_{r}^{(ij)}\left(
  \delta_{ts}-n_{t}^{(ij)}n_{s}^{(ij)}\right) \notag\\
&+&\langle K_{N}\rangle \text{
  }\langle\sum_{a=1,a\neq j}^{N(i)}n_{r}^{(ia)}n_{t}^{(ia)}n_{s}^{(ia)}\rangle\notag\\
  &+&\langle K_{T}\rangle \text{ }\langle \sum_{a=1,a\neq j}^{N(i)}n_{r}^{(ia)}\left( \delta
_{ts}-n_{t}^{(ia)}n_{s}^{(ia)}\right) \rangle,\notag\\
\overline{P}_{rtsg}^{(ij)}&=&\langle K_{N}\rangle \text{ }n_{r}^{(ij)}n_{t}^{(ij)}n_{s}^{(ij)}n_{g}^{(ij)}
+\langle K_{T}\rangle \text{ }n_{r}^{(ij)}\left(
  \delta_{ts}-n_{t}^{(ij)}n_{s}^{(ij)}\right) n_{g}^{(ij)}\notag\\
&+&\langle K_{N}\rangle \text{
  }\langle\sum_{a=1,a\neq
    j}^{N(i)}n_{r}^{(ia)}n_{t}^{(ia)}n_{s}^{(ia)}n_{g}^{(ia)}\rangle\notag\\
&+&\langle K_{T}\rangle \text{ }\langle \sum_{a=1,a\neq j}^{N(i)}n_{r}^{(ia)}\left( \delta
_{ts}-n_{t}^{(ia)}n_{s}^{(ia)}\right)n_{g}^{(ia)} \rangle,
\label{tensors}
\end{eqnarray}
where the average structural sums have been overbarred. These objects are tensors, and in the presence of isotropy, their representation is identical for all contact orientations in the local reference frame of $(\mathbf{n}^{(ij)}, \mathbf{t}^{E(ij)}, \tilde{\mathbf{t}}^{(ij)} )$, where $\tilde{\mathbf{t}}^{(ij)}$ is the unit tangent vector perpendicular to $\mathbf{n}^{(ij)}$ and $\mathbf{t}^{E(ij)}$. Moreover, due to symmetry,
\begin{equation*}
\overline{I}_{rt}^{(ij)}=\overline{I}_{rt}^{(ji)},\;\;\;\overline{J}_{rts}^{(ij)}=-\overline{J}_{rts}^{(ji)},\;\;\;\overline{P}_{rtsg}^{(ij)}=\overline{P}_{rtsg}^{(ji)},
\label{structuralSums2}
\end{equation*}
which reduce equations (\ref{solvSys1}) to 
\begin{equation}
\left[ 
\begin{array}{cc}
\overline{I}^{(ij)}_{rt}\hspace{2cm}\epsilon _{sbc}D\overline{J}^{(ij)}_{brs}
&  \\ 
\epsilon _{sbr}\overline{J}^{(ij)}_{brt}\hspace{0.3cm}\epsilon
_{sbv}\epsilon _{drc}\overline{P}^{(ij)}_{bvdr}& 
\end{array}
\right] \left[ 
\begin{array}{c}
-2\left( \tilde{u}_{t}^{(i)}-\tilde{u}_{t}^{(j)}\right) \\ 
\tilde{\omega} _{c}^{(i)}+\tilde{\omega} _{c}^{(j)}
\end{array}
\right] =\left[ 
\begin{array}{c}
-4D\overline{J}^{(ij)}_{crs}E_{sc} \\ 
-4D^{2}\epsilon _{sbc}\overline{P}^{(ij)}_{bcst}E_{st}
\end{array}
\right],
\label{1FPSolving}
\end{equation}
and equations (\ref{solvSys2}) to a system whose known terms are the same as in (\ref{1FPSolving}), but whose solving matrix:  
\begin{eqnarray}
&&\left[ 
\begin{array}{c}
\overline{I}^{(ij)}_{rt}+\langle K_{rt}\rangle\hspace{2cm}\epsilon _{sbc}D\left( 
\overline{J}^{(ij)}_{brs}+n_{b}^{(ij)}\langle K_{rs}\rangle\right) \\ 
\epsilon _{sbr}\left( \overline{J}^{(ij)}_{brt}+n^{(ij)}_{b}\langle K_{rt}\rangle
\right) \hspace{0.3cm}\epsilon _{sbv}\epsilon _{drc}\left( \overline{P}
^{(ij)}_{bvdr}+n^{(ij)}_{b}\langle K_{vd}\rangle n_{r}\right)
\end{array}
\right]
\label{solvSys}
\end{eqnarray}
still differs from (\ref{1FPSolving}) in the isolated contributions from the chosen contact $ij$. Finally, the averaging process reduces equation (\ref{solvSys3}) to
\begin{eqnarray*}
&&\left[ 
\begin{array}{c}
\overline{I}^{(ij)}_{rt}\hspace{2cm}\epsilon _{sbc}D 
\overline{J}^{(ij)}_{brs} \\ 
\epsilon _{sbc} \overline{J}^{(ij)}_{bct} \hspace{0.3cm}\epsilon _{sbv}\epsilon _{drc}D\overline{P}
^{(ij)}_{bvdr}
\end{array}
\right] \left[ 
\begin{array}{c}
-2\left( \tilde{u}_{t}^{(i)}+\tilde{u}_{t}^{(j)}\right) \\
\tilde{\omega} _{c}^{(i)}-\tilde{\omega} _{c}^{(j)}
\end{array}
\right] =\left[ 
\begin{array}{c}
0 \\ 
0
\end{array}
\right], \label{redSliding1}
\end{eqnarray*}
which differs from the average of equations (\ref{solvSys4}) only in the solving matrix:
\begin{eqnarray*}
&&\left[ 
\begin{array}{c}
\overline{I}^{(ij)}_{rt}-\langle K_{rt}\rangle\hspace{2cm}\epsilon _{sbc}D 
\left(\overline{J}^{(ij)}_{brs}-n_{b}^{(ij)}\langle K_{rs}\rangle\right) \\ 
\epsilon _{sbc} \left( \overline{J}^{(ij)}_{bct}-n^{(ij)}_{b}\langle K_{ct}\rangle
\right) \hspace{0.3cm}\epsilon _{sbv}\epsilon _{drc}D\left(\overline{P}
^{(ij)}_{bvdr}-n^{(ij)}_{b}\langle K_{vd}\rangle n_{r}\right)
\end{array}
\right].
\label{redSliding2}
\end{eqnarray*}
With respect to (\ref{solvSys1}), (\ref{solvSys2}), (\ref{solvSys3}) and (\ref{solvSys4}), the average contribution from the chosen contact emerges more neatly. Equations (\ref{1FPSolving}) and (\ref{solvSys}) are sufficient to determine the elastic moduli, as they can be
solved for the
average relative displacements that do contribute to the contact forces. As $\tilde{u}_{t}^{(i)}+\tilde{u}_{t}^{(j)}$ and $\omega _{b}^{(i)}-\omega
_{b}^{(j)}$ do not contribute to (\ref{1FPSolving}) and (\ref{solvSys}), the average total force and torque the neighbors exert on the pair as a result of its rolling and of its rigid motion are zero in these approximation schemes. The reader is referred to (Bagi and Kuhn, 2004) and (Kuhn and Bagi, 2004) for a review of
recent results about rolling in granular media, which is deferred for future work.

The remaining step towards a continuum equivalent description would consist in the search for an appropriate analytical formulation of (\ref{1FPSolving}) and (\ref{solvSys}). As far as the PF scheme is concerned, two formulations are proposed, in Jenkins et al (2005) for spheres and in Agnolin et al. (2005) for disks. However, those both consider the isolated contributions from the
contact $ij$ in (\ref{solvSys}) of the second order and neglect it, and thus correspond to the 1FP case. In more general terms, the key element to the analytical formulation is the choice of appropriate distribution functions for
the contact orientation, which allow to transform the structural sums
into integrals. As already noticed by Jenkins (1997), such a choice cannot stem from the focus on the individual particle alone, as in this case symmetry in the contact distribution would make the third order tensor
$J_{rst}$ in (\ref{tensors}) be zero. If the third order tensor were zero, from (\ref{1FPSolving}) the center displacements would be zero as well, and rotations would balance alone the effect of a biaxial
load, in contradiction to both experiments and numerical simulations. Therefore, adequate statistical representations of the average structure need to incorporate considerations of impenetrability between neighboring particles. On this base, Jenkins (1997) concluded on the inadequacy of the 1FP approach and later on introduced the PF method, while here it has been proven that the 1FP approach is already effective in capturing the fluctuations, at the condition that the focus be extended to the pair.

\section{Conclusions and perspectives}

Four numerical isotropic assemblies of frictional spheres have
been created by employing different preparation procedures. Of these, the first one is the
classic frictionless isotropic compression; two procedures aim
to reproduce compaction by vibration and lubrication, respectively, while the fourth one is a frictional
compression. The coordination number of the four samples obtained in this way is found to strongly vary with the preparation procedure, and their elastic moduli depend much more 
on coordination number than on density. The initial states of the
first three samples
all match in pressure and density that of assemblies of glass beads used
during a series of wave propagation experiments, but as far as the elastic moduli are concerned, only those of the low-coordinated sample produced by the protocol aimed at mimicking vibration match the
experimental results satisfactorily at pressures lower than $1$ $MPa$. Conversely, samples created by means of an initial isotropic compression, the scientific literature usually
focuses on, are far too stiff in this range of pressure.

The interest of Continuum Mechanics research is in the prediction of the mechanical properties, which requires that of the relative displacement between contacting grains. In these respects, the easiest scenario possible is that of relative displacements driven by the average strain. Such an hypothesis results in poor estimates of the shear modulus, which indicates the necessity of incorporating the displacement fluctuations from the average strain. We have found that the fluctuations chiefly depend on coordination number. Locally, the fluctuations
vary strongly, as they ensure the balance of force and torque at the grain
scale, but only their average over contacts with the same orientation affects the
macroscopic behavior. This average is determined by the
average geometry and the average contact stiffness. In more detail, the average normal and tangential component of the
contact displacements induced by the fluctuations are proportional, in turn, to the normal and the tangential relative
displacement induced by the average strain. Along the tangential
direction, the relaxation with respect to the average strain assumption is generally
entirely due to rotations; the tangential fluctuations of the centers give a much smaller
contribution, and mainly originate in the requirement of local equilibrium.

Displacement fluctuations consistent with these observations result from the
1FP and the PF approach, which have been promisingly employed in the past to predict the moduli of frictionless spheres and frictional disks, and which have been tested numerically on our assemblies of frictional spheres. In this way, the bound has been identified which can be attained by the corresponding analytical solutions for frictional spheres. Both approaches are based on the
equilibrium of a small subassembly made, in turn, of a particle and a pair
of contacting grains with their first neighbors. The chosen particle or pair alone is
allowed to fluctuate, while the neighborhood move in accordance with the
average strain. Both approaches result in a remarkable improvement of the
estimate of the shear modulus with
respect to the average strain assumption, meaning that the fluctuations which mostly affect the macroscopic behavior correlate over a short lenght. The estimate is satisfying in the case of samples with
large coordination number, and optimal predictions for samples with a smaller
coordination number are likely to result from
just incorporating the fluctuations of the first neighbors. Tangential fluctuations are
rather satisfactorily captured, and the residual discrepancy between
predicted and effective shear modulus is mostly due to the underestimate of
radial fluctuations.

\section *{\protect\normalsize Acknowledgments}

{\normalsize The authors acknowledge financial support of the first author,
during her stay at the LCPC, from the ''AREA, Consorzio Area di Ricerca,
Trieste, Italy'' and the Region Ile-de-France. This work benefitted from
discussion with Prof. J.T. Jenkins, Cornell University, USA, Dr. L. La
Ragione, Politecnico di Bari (Italy) and Dr.N.P.
Kruyt, Twente University, The Netherlands. }

\end{document}